\documentclass[11pt,a4paper]{article}
\usepackage{jheppub}
\usepackage[utf8x]{inputenc}
\usepackage{amsmath}
\usepackage{amssymb}
\usepackage{float}
\usepackage{array}
\usepackage{mathtools}
\usepackage{mathrsfs} 
\numberwithin{equation}{section}
\usepackage{marvosym}

\usepackage{tikz}
\usetikzlibrary{arrows}
\tikzstyle{emptybox} = [rectangle, rounded corners, minimum width=.5cm, minimum height=.5cm,text centered, draw=white, fill={white}]
\tikzstyle{arrow} = [thin,->,>=angle 60]
\tikzstyle{arrowb} = [thin,<-,>=angle 60]
\tikzstyle{bigarrow} = [thick,<->,>=angle 60]

\usepackage{tikz-cd}

\title{\boldmath Permutations of Massive Vacua}
 \author[a]{Antoine Bourget} 
 \author[b]{and Jan Troost}
  \affiliation[a]{Department of Physics, Universidad de Oviedo \\
Avenida Calvo Sotelo 18, 33007 Oviedo, Spain}
 \affiliation[b]{Laboratoire de Physique Théorique de l'École Normale Supérieure \\
 CNRS, PSL Research University, Sorbonne Universités, 75005 Paris, France}

\emailAdd{bourgetantoine@uniovi.es}
\emailAdd{troost@lpt.ens.fr}

\abstract{We discuss the permutation group $G$ of massive vacua of four-dimensional gauge theories with ${\cal N}=1$ supersymmetry that arises upon tracing loops in the space of couplings. We concentrate on superconformal ${\cal N}=4$ and ${\cal N}=2$ theories with ${\cal N}=1$ supersymmetry preserving mass deformations. 
The permutation group $G$ of massive vacua is the Galois group of characteristic polynomials for the vacuum expectation values of chiral observables. 
We provide various techniques to effectively compute characteristic polynomials in given theories, and we deduce the existence of varying symmetry breaking patterns of the duality group depending on the gauge algebra and matter content of the theory. Our examples  give rise to interesting field extensions of spaces of modular forms. }

\begin{document} 
\maketitle
\flushbottom

\section{Instigation}
\label{section1}

Four-dimensional gauge theories play a paramount role in our description of nature. While we understand the phase structure of gauge theories reasonably well, we have an even better control over the subset of theories with supersymmetry. The minimal amount of supersymmetry in four dimensional Poincar\'e invariant theories is four supercharges, or ${\cal N}=1$. We comprehend a lot of the vacuum structure of ${\cal N}=1$ theories, including the number of vacua for the pure ${\cal N}=1$ super Yang-Mills theory, methods to solve for the superpotential in any given vacuum, as well as the systematics of the chiral ring of ${\cal N}=1$ theories (see e.g. \cite{Witten:2000nv}, \cite{Dijkgraaf:2002dh} and \cite{Cachazo:2003yc} respectively). While the theory of the chiral sector of ${\cal N}=1$ theories is well-grounded, concrete answers about the vacuum structure or the chiral ring of particular theories can still be hard to obtain. In this paper, we develop further tools to characterize the vacua.

We consider quantum field theories that can be defined using discrete data (for instance, the gauge group and the field content) as well as parameters that take values in a continuous space (e.g. the coupling constants). We call the space of parameters $\bar{C}$, which is a connected space by definition. In a regime where a Lagrangian description is available, the space $\bar{C}$ can be identified with the parameters needed to define this Lagrangian, such as gauge coupling constants and masses. If the theory permits duality transformations, those appear as discrete quotients and $\bar{C}$ is an orbifold. 

We focus on $\mathcal{N}=1$ supersymmetric gauge theories in four dimensions. In such theories, an efficient probe of the space $\bar{C}$ is the vacuum structure, which is a fiber bundle with base $\bar{C}$. More precisely, we expect to find at each point of $\bar{C}$ a finite number of massive isolated vacua, along with massless vacua that form continuous branches. Let us call $V_{\textrm{massive}}$ the space of massive vacua: it is a fiber bundle where the fiber is finite or in other words, $V_{\textrm{massive}}$ is an $n$-fold cover of $\bar{C}$, where $n$ is the number of massive vacua. However, there might be points in parameter space where some or all vacua  become indistinguishable -- for instance at the zero coupling point in pure $\mathcal{N}=1$ Yang-Mills theory. We assume that these points are isolated.
In order to characterize completely both the parameter space and the bundle of massive vacua, we will restrict our attention to the subspace $C \subset \bar{C}$ where the massive vacua are distinguishable, i.e. the space $C$ is the parameter space $\bar{C}$ with points where vacua merge excluded. 

As a consequence, when moving along a non-contractible cycle of $C$ the $n$ vacua are permuted in a well defined way. This means that the cover $V_{\textrm{massive}} \rightarrow C$ can be described through a map 
\begin{equation}
\label{mapPi1}
    \sigma : \pi_1 (C) \rightarrow \mathfrak{S}_n \,  ,
\end{equation}
where $\mathfrak{S}_n$ is the permutation group of $n$ elements acting on the massive vacua of the theory. The full parameter space $\bar{C}$ is in general an orbifold characterized by the fact that its fundamental group $\pi_1 (\bar{C})$ is a maximal quotient of $\pi_1 (C)$ such that $\sigma$ passes to the quotient. In other words, the fundamental group of $\bar{C}$ is the smallest subgroup of $\pi_1(C)$ whose image under $\sigma$ is all of $\mathrm{Im} \, \sigma$, and the following diagram commutes
\begin{equation}
\begin{tikzcd}
  \pi_1 (C)  \arrow{rd}{\sigma} \arrow{d} &  \\
\pi_1 (\bar{C}) \arrow{r}{\bar{\sigma}} & \mathfrak{S}_n
\label{sigmabar}
\end{tikzcd}
\end{equation}
One goal of this paper is to develop tools that allow to understand the map $\sigma$, which in turn gives precious information about the two objects that it relates: 
\begin{itemize}
    \item On the $\mathfrak{S}_n$ permutation group side, we learn how the vacua are transformed when the parameters are varied. For reasons that will become clear in the following, we call  the image of $\sigma$ the   
    \emph{Galois group}
    \begin{equation}
    G = \mathrm{Im} \, \sigma = \mathrm{Im} \, \bar{\sigma} \, . 
    \end{equation}
    Amongst other information, this provides knowledge on which semi-classically distinct vacua can be continuously connected in the quantum theory.
    \item On the $\pi_1$ side, we learn about the structure of the space $\bar{C}$. The homotopy subgroup $\Gamma_{\textrm{Galois}}$ 
    is identified with the kernel of $\sigma$, 
    \begin{equation}
    \Gamma_{\textrm{Galois}}
    = \mathrm{Ker} \, \sigma \, . 
    \end{equation}
\end{itemize}
Our main tool of analysis is the characteristic polynomial associated to an operator $\mathcal{O}$, defined by 
\begin{equation}
\label{generalDefinitionPoly}
  P_{\mathcal{O}} (X) = \prod_{i =1}^n (X- \langle \mathcal{O} \rangle_i) \, , 
\end{equation}
where $\langle \mathcal{O} \rangle_i$ is the expectation value in the vacuum $i$, $X$ is a formal variable and we multiply
over the set of $n$ vacua. Although $\langle \mathcal{O} \rangle_i$ is multi-valued on $C$, the characteristic polynomial is well defined and single-valued on $C$, and if the operator $\mathcal{O}$ is \emph{generic}, the Galois group associated to the polynomial\footnote{See e.g. \cite{Baker,Stewart} for pedagogical introductions.} characterizes completely the way the vacua are permuted \cite{Ferrari:2008rz,Ferrari:2009zh}: 
\begin{equation}
\mathrm{Gal}\left(P_{\mathcal{O}}\right) = G = \mathrm{Im} \, \sigma \, . 
\end{equation} 
We say that $\mathcal{O}$ is \emph{generic} if its Galois group is maximal, in the sense that for any operator $\mathcal{O}'$, we have that the Galois group of $P(\mathcal{O})$ satisfies $\mathrm{Gal}\left(P_{\mathcal{O}'}\right) \subseteq \mathrm{Gal}\left(P_{\mathcal{O}}\right)$. Determining the Galois group of a polynomial or of a field extension is generically a difficult task;  we will show on various examples how to overcome the difficulties by combining algebraic facts, group theory databases like \cite{finitegroups,Pfeiffer} and the software Sage \cite{Sage}.

The analysis of the Galois group of the characteristic polynomial will serve to determine the breaking pattern of the symmetries of the theory under massive deformation. Specifically, in the mass deformed ${\cal N}=2$ superconformal theories on which we concentrate, we will characterize the breaking of modular invariance in the vacua in this manner. 
Moreover, the study of the characteristic polynomial of the superpotential will also lead to further insight into the vacuum expectation values of the superpotential. The latter are useful for instance in determining domain wall tensions, or the coupling dependence of the glueball expectation value.

\subsection*{Plan of the paper}

Our ideas are illustrated in examples of growing complexity. The whole of sections \ref{SectionOnestar}, \ref{SectionSun} and \ref{SectionSo8} are dedicated to  mass deformed
${\cal N}=4$ theories. In section \ref{SectionOnestar}, we demonstrate that the characteristic polynomial of the extremal superpotential values in this theory is a polynomial with coefficients in the graded ring of modular forms. We analyze how the roots, i.e. the extremal superpotential values, extend the field of functions on modular curves, and generate the corresponding Galois group. The analysis of the characteristic polynomial shows that there are two possibilities for the fate of modular invariance in the vacua: 
\begin{itemize}
    \item There can be a residual modular invariance, that is characterized by the fact that $\Gamma_{\textrm{Galois}}$ is a congruence subgroup of $\mathrm{PSL}(2,\mathbb{Z})$. The theories with gauge algebra $\mathfrak{su}(N)$ fall in this category, as shown in section \ref{SectionSun}. We show that in general the Galois group is a strict subgroup of the permutation group, and we compute it explicitly in low-rank examples. 
    \item Modular invariance can be completely broken. In this case the roots of the polynomial with coefficients in the ring of modular forms provide an extension which goes beyond the theory of modular forms for congruence subgroups and includes functions with singularities inside the fundamental domain. The theory with gauge algebra $\mathfrak{so}(8)$ has this property. It is analyzed in section \ref{SectionSo8}. 
\end{itemize}

In section \ref{SectionQCD} we illustrate how the concepts we covered in the case of the  mass deformed ${\cal N}=4$ theory generalize to ${\cal N}=2$ superconformal QCD broken by mass terms to an ${\cal N}=1$ supersymmetric theory. In the process, we provide a detailed analysis of how the duality group acts in the mass deformed theory, and how the permutation group supersedes the duality group in describing the set of massive vacua. We prove that superconformal QCD belongs to the class of theories where modular invariance is completely broken in a given vacuum.

Our detailed discussion  builds in part on previous work,
in which we analyzed the vacuum structure of ${\cal N}=4$ supersymmetric Yang-Mills theory with mass deformations for the three adjoint chiral multiplets. This ${\cal N}=1^\ast$ theory has a  rich structure of massive vacua \cite{Dorey:1999sj,Polchinski:2000uf}. In the work \cite{Bourget:2015cza}, we used a combination of numerical analysis of the associated elliptic integrable system \cite{Donagi:1995cf,Dorey:1999sj,Kumar:2001iu}, as well as expectations from gauge theory to amass a treasure trove of data, which required further interpretation on several points. In \cite{Bourget:2015lua}, we provided a systematic and simplified count of the number of massive vacua of all ${\cal N}=1^\ast$ theories on $\mathbb{R}^4$. In \cite{Bourget:2015upj}, we analyzed  the count of vacua upon compactifying the theory on a circle, and the analytic description of massless branches of vacua. In \cite{Bourget:2016yhy} we showed how the count of massive vacua for the compactified theory, and its invariance under duality, leads to intriguing number theoretic identities. 

Using the general
picture described in this paper we further our understanding of the wealth of data we excavated in \cite{Bourget:2015cza}. We acquire a deeper algebraic and analytic understanding of both the extremal superpotential values in the massive vacua as well as the duality symmetry breaking patterns by embedding our findings in Galois theory.

\section{\texorpdfstring{Mass Deformed ${\cal N}=4$ Super Yang-Mills Theories}{}}
\label{SectionOnestar}

In this section, we study the ${\cal N}=4$ super Yang-Mills theory deformed by three mass terms for the three ${\cal N}=1$ chiral multiplets in the adjoint. In this deformed theory, an operator that distinguishes all the vacua at generic coupling is the superpotential $W$. The superpotential will therefore serve as
our generic operator $\mathcal{O}$, in the sense of section \ref{section1}. 
We first recall a few salient features of the ${\cal N}=1^\ast$ theory and its vacua. Then we write down the characteristic polynomial of the superpotential and explain how it  diagnoses the way in which the modular invariance of the ${\cal N}=4$ theory is partially broken in the vacua. The theory is then illustrated in the simplest case of the $\mathfrak{su}(2)$ theory. We  provide example applications as well as remarks on the relation to the vacuum expectation values of the full chiral ring. In section \ref{SectionSun} we will further exemplify our analysis with  the $\mathfrak{su}(N)$ theories with $N \ge 3$.

\subsection{\texorpdfstring{Duality in the ${\cal N}=4$ Theory and Permutations of ${\cal N}=1^\ast$ Vacua}{}}
\label{generalitiesPolynomials}
In this subsection, we identify the parameter space, the characteristic polynomial, the relevant Galois theory, and the reduced modular covariance of the theory in a given vacuum.

\subsubsection{The Parameter Space}

We recall that the  ${\cal N}=4$ super Yang-Mills theories with  gauge algebra $\mathfrak{g}$ take on various guises, depending on the global choice of gauge group, and the spectrum of line operators \cite{Aharony:2013hda}. The implications of these global choices are commented on in the case of the Lie algebra $\mathfrak{g} = \mathfrak{su}(N)$ in the appendix, section \ref{appendixTheoriessuN}. In the following, we will focus on local physics on the topologically trivial space $\mathbb{R}^{1,3}$, which depends only on the choice of Lie gauge algebra. We will correspondingly refer to \emph{the} ${\cal N}=4$ theory with gauge algebra $\mathfrak{g}$. 

When the algebra $\mathfrak{g}$ is simply laced, which will be the case throughout this article, the ${\cal N}=4$ theory is invariant under the action of the modular group $\mathrm{PSL}(2,\mathbb{Z})$ acting on the complexified coupling constant $\tau$. The parameter space can then be identified with the quotient 
\begin{equation}
\label{parameterspaceN4}
\bar{C}_{{\cal N}=4} = \mathrm{PSL}(2,\mathbb{Z}) \backslash \mathfrak{H}
\end{equation}
of the upper half-plane $\mathfrak{H}$ by this group. 
The half-plane $\mathfrak{H}$ contains two equivalence classes of fixed points for the action of $\mathrm{PSL}(2,\mathbb{Z})$ and as a result, the space $\bar{C}_{{\cal N}=4}$ is a non-trivial orbifold. It can be seen as a space with the topology of a plane with two orbifold points of order two and three respectively. Using the appropriate notion of loops in an orbifold, we can define the fundamental group of $\bar{C}_{{\cal N}=4}$ and find
\begin{equation}
   \pi_1 (\bar{C}_{{\cal N}=4}) = \mathbb{Z}_2 * \mathbb{Z}_3  = \mathrm{PSL}(2,\mathbb{Z}) \, ,
\end{equation}
where the symbol $*$ denotes the free product between groups. 

When we deform ${\cal N}=4$ on $\mathbb{R}^{1,3}$ by three mass terms for the adjoint chiral multiplets, we obtain the ${\cal N}=1^\ast$ theory. This theory has a finite number $n$ of massive vacua, which can be characterized by the expectation value of the superpotential $W$. Since the vacua are identified with the extrema of the superpotential, the latter is a good generic operator whose characteristic polynomial allows to probe the vacuum structure. 

The space of massive vacua is a bundle $V_{\textrm{massive}}$ over the parameter space with finite fiber. The parameter space includes the three masses, in addition to the space $\bar{C}_{{\cal N}=1^\ast} (\mathfrak{g})$ corresponding to the complexified coupling constant. As long as the masses remain non-zero, they can be rescaled without changing the vacuum structure, and therefore they factorize in the bundle $V_{\textrm{massive}} \rightarrow \bar{C}_{{\cal N}=1^\ast} (\mathfrak{g})$. As a consequence, a minimal hypothesis is that even after the mass deformation, the parameter space of the ${\cal N}=1^\ast$ theory is a one complex dimensional manifold $\bar{C}_{{\cal N}=1^\ast} (\mathfrak{g})$. When no extra singularity is introduced by the mass deformation, this manifold is equal to the space (\ref{parameterspaceN4}). We will loop back to this hypothesis and discuss that Occam's razor may well cut too deep.

\subsubsection{The Characteristic Polynomial}

We analyze the structure of the bundle of massive vacua over the coupling space $C_{{\cal N}=1^\ast} (\mathfrak{g})$ using the values of the superpotential at its $n$ massive extrema. For the $\mathfrak{su}(N)$ theories, the superpotential extrema $W_i$ in the massive vacua are known explicitly \cite{Dorey:1999sj,Polchinski:2000uf,Ritz:2006ji}, while for other gauge algebras only partial results are available \cite{Bourget:2015cza}. We define the characteristic superpotential polynomial $P_{\tau}$ which has as roots the values of the superpotential in the massive vacua at gauge coupling $\tau \in C_{{\cal N}=1^\ast} (\mathfrak{g})$, and coefficient one for the maximal degree term:
\begin{eqnarray}
\label{characteristicPoly}
P_{\tau} (W) &=& \prod_{i =1}^n (W-W_i (\tau)) \, .
\end{eqnarray}
According to equation (\ref{mapPi1}), if there is no extra singularity, the superpotentials $W_i (\tau)$ constitute a vector-valued modular form of weight two for the modular group $\mathrm{PSL}(2,\mathbb{Z})$ where all the elements of this group are represented by permutation matrices. In any case, the coefficients of (\ref{characteristicPoly}) are symmetric functions of the roots, and therefore are modular forms of $\mathrm{PSL}(2,\mathbb{Z})$ with weight equal to twice their degree in the values $W_i$, which means that they belong to the graded ring $\mathfrak{m}$ of polynomials in $E_4$ and $E_6$ with coefficients in $\mathbb{C}$
\begin{equation}
\label{ringMF}
    \mathfrak{m} = \mathbb{C}[E_4 , E_6] \, .
\end{equation}
Here and in the following, the functions $E_{2n}$ for $n \geq 1$ are the Eisenstein series of weight $2n$ with unit leading coefficient, and Fourier expansion
\begin{equation}
E_2 (\tau) = 1-24  \sum\limits_{n=1}^{\infty} \frac{n   q^n}{1-q^n} \, , \qquad 
E_4 (\tau) = 1+240 \sum\limits_{n=1}^{\infty} \frac{n^3 q^n}{1-q^n} \, , \qquad 
E_6 (\tau) = 1-504 \sum\limits_{n=1}^{\infty} \frac{n^5 q^n}{1-q^n} \, , \qquad
\end{equation}
with $q=e^{2 \pi i \tau}$. The graded dimension of the ring $\mathfrak{m}$ is encoded in the Hilbert series 
\begin{equation}
\label{gradeddimm}
   \mathrm{dim} \, \mathfrak{m} :=  \sum\limits_{k=0}^{\infty} t^k \mathrm{dim} \, \mathcal{M}_k (\mathrm{PSL}(2,\mathbb{Z})) = \frac{1}{(1-t^4)(1-t^6)} \, . 
\end{equation}
We assign a modular weight two to the formal variable $W$ such that the whole characteristic polynomial has a modular weight equal to twice its degree. It will also be useful to introduce another degree $n$ polynomial $\tilde{P}_{\tau} (Z)$, where $Z = \frac{E_4}{E_6} W$ is a formal variable of weight zero, defined by 
\begin{equation}
\label{characteristicPolyWeight0}
   \tilde{P}_{\tau} (Z) = \left( \frac{E_4}{E_6} \right)^n P_{\tau} \left( \frac{E_6}{E_4} Z \right) \, . 
\end{equation}
The coefficients of this polynomial are meromorphic functions on the space (\ref{parameterspaceN4}), which can be written as quotients of polynomials in the modular Klein invariant 
\begin{equation}
   j(\tau) = \frac{1728 E_4^3(\tau)}{E_4^3(\tau) - E_6^2(\tau)} \, , 
\end{equation}
or in other words $\tilde{P}_{\tau} \in \mathfrak{k}[Z]$ where $\mathfrak{k} = \mathbb{C} (j)$.

\subsubsection{The Galois Correspondence}

Consider an $\mathcal{N}=1^\ast$ theory with a simply laced Lie gauge algebra $\mathfrak{g}$, with its characteristic polynomial $P_{\tau} \in \mathfrak{m}[W]$ constructed in (\ref{characteristicPoly}). The degree $n$ of this polynomial is equal to the number of massive vacua, and its roots $W_i$ are pairwise distinct.\footnote{It is possible that the superpotentials in two different vacua coincide for some particular value of $\tau$, as it happens for instance in the limit $\tau \rightarrow i \infty$. However, the $W_i$ considered as functions of $\tau$ are pairwise distinct. } 
The polynomial defines a map 
\begin{eqnarray}
\label{mapP}
   P &:& C_{{\cal N}=1^\ast} (\mathfrak{g}) \rightarrow \mathbb{C}[W] \nonumber \\
   & & \tau \mapsto P_\tau \, . 
\end{eqnarray}
This map provides an explicit realization of the function $\sigma$ defined in (\ref{mapPi1}). Indeed, along a loop drawn in $C_{{\cal N}=1^\ast} (\mathfrak{g})$, on which by definition the roots of the polynomial are always distinct, we can continuously follow these roots and compute their permutation. This leads us to introduce the discriminant of the characteristic polynomial
\begin{equation}
    D(\tau) = \mathrm{Discriminant} (P_\tau) = \prod\limits_{i<j} \left(W_i(\tau) - W_j(\tau)\right)^2 \in \mathfrak{m} - \{0\} \, . 
\end{equation}
We expect to have permutations of roots when looping around the orbifold points, and around points where two or more roots coincide, which appear as zeros of the discriminant. A zero with a non-trivial monodromy will be called a \emph{non-trivial zero}. In particular, any end of a branch cut of $\sqrt{D(\tau)}$ is a non-trivial zero. 

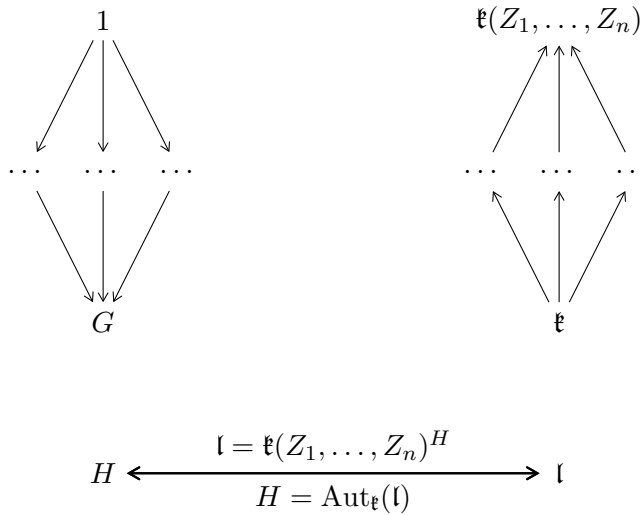
\begin{figure}
\begin{center}
\begin{tikzpicture}[node distance=2cm]
\node (1)[emptybox] at (0,0) {$G$};
\node (2)[emptybox, above of=1, xshift=-1cm] {$\dots $};
\node (3)[emptybox, above of=1] {$\dots $};
\node (4)[emptybox, above of=1, xshift=+1cm] {$\dots $};
\node (5)[emptybox, above of=3] {$1$};
\draw [arrow] (2) -- (1);
\draw [arrow] (3) -- (1);
\draw [arrow] (4) -- (1);
\draw [arrow] (5) -- (2);
\draw [arrow] (5) -- (3);
\draw [arrow] (5) -- (4);
\node (6)[emptybox] at (6,0) {$\mathfrak{k}$};
\node (7)[emptybox, above of=6, xshift=-1cm] {$\dots $};
\node (8)[emptybox, above of=6] {$\dots $};
\node (9)[emptybox, above of=6, xshift=+1cm] {$\dots $};
\node (10)[emptybox, above of=8] {$\mathfrak{k}(Z_1 ,\dots , Z_n)$};
\draw [arrowb] (7) -- (6);
\draw [arrowb] (8) -- (6);
\draw [arrowb] (9) -- (6);
\draw [arrowb] (10) -- (7);
\draw [arrowb] (10) -- (8);
\draw [arrowb] (10) -- (9);
\node (21)[emptybox] at (0,-2) {$H$};
\node (22)[emptybox] at (6,-2) {$\mathfrak{l}$};
\draw [bigarrow] (21) -- node[anchor=north] {$H = \mathrm{Aut}_{\mathfrak{k}}(\mathfrak{l})$}(22);
\draw [bigarrow] (22) -- node[anchor=south] {$\mathfrak{l} = \mathfrak{k}(Z_1 ,\dots , Z_n)^{H}$}(21);
\end{tikzpicture}
\caption{Schematic Galois correspondence }
\label{GaloisCorrespondence}
\end{center}
\end{figure}

The field extension $\mathfrak{k}(Z_1 , \dots , Z_n)$ is by definition a Galois extension,\footnote{We recall that a Galois extension is characterized by the fact that it is the splitting field of some separable polynomial with coefficients in the base field. A polynomial is called separable if its roots in any algebraic closure of the base field are distinct. } and the fundamental theorem of Galois theory states that there exists a finite group
\begin{equation}
   G = \mathrm{Gal} \left( \mathfrak{k}(Z_1 , \dots , Z_n) / \mathfrak{k} \right) 
\end{equation}
such that the subgroup structure of $G$ coincides with the field extensions that sit between $\mathfrak{k}$ and $\mathfrak{k}(Z_1 , \dots , Z_n)$, as illustrated in figure \ref{GaloisCorrespondence}. This gives us a cross-fertilizing relation between global aspects of the geometry of the parameter space, characterized by the Galois group, and local data about individual vacua. 

Note that if $\sqrt{D(\tau)}$ has branch cuts, or more generally non-trivial zeros, the modular transformations like $T$ or $S$ become path-dependent. Correspondingly, the superpotential values in the multiplets of vacua that contain roots of a characteristic polynomial with such a discriminant no longer exhibit 
modular covariance. This situation will be illustrated in section \ref{SectionSo8}. On the other hand, if there is no non-trivial zero, modular transformations can be performed, and modular covariance is broken in any given vacuum to a reduced modular covariance that can be studied using Galois theory, as described in the next paragraph.  

\subsubsection{Reduced Modular Covariance}
\label{sectionTripleGalois}

In this subsection, we assume that $D(\tau)$ has no non-trivial zero. From the polynomial we can then always construct a duality diagram  which represents the action of the generators $S$ and $T$ of the modular group. In other words, the vector $(W_1 , \dots , W_n)$ is a weight two vector-valued modular form. The fact that $D(\tau)$ has no non-trivial zero assures that this action is well-defined and independent from the path followed to compute it. There is a one-to-one correspondence between irreducible factors of
the polynomial $P$ over the ring $\mathfrak{m}$ and connected components of the duality diagram, and for simplicity we will focus in this section on an irreducible factor. 

We then have a transitive permutation representation of $\mathrm{PSL}(2,\mathbb{Z})$, and this defines uniquely a conjugacy class of subgroups of $\mathrm{PSL}(2,\mathbb{Z})$. There is an algorithm \cite{hsu} that allows to know whether this conjugacy class contains a congruence subgroup or not.\footnote{Definitions and useful facts about congruence subgroups are collected in appendix \ref{appendixCongSub}.} In the following, we assume the conjugacy class does contain a congruence subgroup that we call $\Gamma_{\mathrm{roots}}$.\footnote{\label{footnoteAssumption}For practical purposes, this assumption is not needed. In any particular situation, one simply computes the group $\Gamma_{\mathrm{roots}}$ assuming that it is a congruence subgroup, and then checks that it indeed leaves one root invariant. } This notation is justified by the fact that the roots of the characteristic polynomial are in bijection with coset representatives of $\Gamma_{\mathrm{roots}}$. Congruence subgroups are fairly well known, and an exhaustive list has been computed for genus lower than 24, see \cite{cummins}. In order to identify the congruence subgroup, one first reads off from the permutation diagram 
\begin{itemize}
    \item the number $c_2$ of roots fixed by $S$, 
    \item the number $c_3$ of roots fixed by $ST^{-1}$,
    \item the partition $\rho$ of $n$ given by the $T$-cycles structure, and in particular the number $c_{\infty}$ of $T$-cycles. 
\end{itemize}
The genus of the group is then given by 
\begin{equation}
\label{genusFormula}
   g = 1 + \frac{n}{12} - \frac{c_{\infty}}{2} - \frac{c_3}{3} - \frac{c_2}{4} \, . 
\end{equation}
This set of data entirely characterizes the group $\Gamma_{\mathrm{roots}}$.\footnote{In particular, we remind the reader that the graded dimension of the ring of modular forms for any congruence subgroup $\Gamma$ that contains the matrix $-I$ is given by \cite{diamond}
\begin{equation}
\label{dimringMF}
   H_{\Gamma}(t ) = \mathrm{dim} \, \mathcal{M} (\Gamma) = 1+ \sum\limits_{k=1}^{\infty} t^{2k} \left( (2k-1)(g-1) + \left \lfloor{\frac{k}{2}}\right \rfloor c_2 + \left \lfloor{\frac{2k}{3}}\right \rfloor c_3 + k c_{\infty} \right) \, . 
\end{equation} 
If $\Gamma$ does not contain $-I$, the above formula still gives the correct contribution for even powers of $t$, but there is an additional contribution from odd powers of $t$ for which we found no general formula. }
If $g \leq 24$, it can  be found in the tables of \cite{cummins} using the above data and the $T$-multiplet structure. The conjugates of $\Gamma_{\mathrm{roots}}$ are the stabilizers of the individual roots of the characteristic polynomial. Their intersection is a congruence\footnote{The group $\Gamma (N)$ is normal in $\mathrm{PSL}(2,\mathbb{Z})$ as it is the kernel of the reduction modulo $N$. Then if $\Gamma(N) \subset \Gamma_{\mathrm{roots}}$, the group $\Gamma(N)$ is also contained in all the conjugates of $\Gamma_{\mathrm{roots}}$, and in their intersection. } subgroup that is equal to $\Gamma_{\mathrm{Galois}}$. 

\begin{figure}
\begin{center}
\begin{tikzpicture}[node distance=2cm]
\node (1)[emptybox] at (0,0) {$G$};
\node (2)[emptybox, above of=1, xshift=-1cm] {$\dots $};
\node (3)[emptybox, above of=1] {$\dots $};
\node (4)[emptybox, above of=1, xshift=+1cm] {$\dots $};
\node (5)[emptybox, above of=3] {$1$};
\draw [arrow] (2) -- (1);
\draw [arrow] (3) -- (1);
\draw [arrow] (4) -- (1);
\draw [arrow] (5) -- (2);
\draw [arrow] (5) -- (3);
\draw [arrow] (5) -- (4);
\node (6)[emptybox] at (5,0) {$\mathfrak{k}$};
\node (7)[emptybox, above of=6, xshift=-1cm] {$\dots $};
\node (8)[emptybox, above of=6] {$\dots $};
\node (9)[emptybox, above of=6, xshift=+1cm] {$\dots $};
\node (10)[emptybox, above of=8] {$\mathfrak{k}(Z_1 ,\dots , Z_n)$};
\draw [arrowb] (7) -- (6);
\draw [arrowb] (8) -- (6);
\draw [arrowb] (9) -- (6);
\draw [arrowb] (10) -- (7);
\draw [arrowb] (10) -- (8);
\draw [arrowb] (10) -- (9);
\node (11)[emptybox] at (10,0) {$\mathrm{PSL}(2,\mathbb{Z})$};
\node (12)[emptybox, above of=11, xshift=-1cm] {$\dots $};
\node (13)[emptybox, above of=11] {$\dots $};
\node (14)[emptybox, above of=11, xshift=+1cm] {$\dots $};
\node (15)[emptybox, above of=13] {$\Gamma_{\mathrm{Galois}}$};
\draw [arrowb] (12) -- (11);
\draw [arrowb] (13) -- (11);
\draw [arrowb] (14) -- (11);
\draw [arrowb] (15) -- (12);
\draw [arrowb] (15) -- (13);
\draw [arrowb] (15) -- (14);
\draw [bigarrow] (4) -- (7);
\draw [bigarrow] (9) -- (12);
\node (21)[emptybox] at (0,-2) {$H$};
\node (22)[emptybox] at (5,-2) {$\mathfrak{l}$};
\node (23)[emptybox] at (10,-2) {$\Gamma$};
\draw [bigarrow] (21) -- node[anchor=north] {$H = \mathrm{Aut}_{\mathfrak{k}}(\mathfrak{l})$}(22);
\draw [bigarrow] (22) -- node[anchor=south] {$\mathfrak{l} = \mathfrak{k}(Z_1,\dots )^{H}$}(21);
\draw [bigarrow] (22) -- node[anchor=north] {$\mathfrak{l} = \mathbb{C}(\mathbf{X}(\Gamma))$}(23);
\draw [bigarrow] (23) -- node[anchor=south] {$\Gamma = \mathrm{PSL}(2,\mathbb{Z})^{\mathfrak{l}}$}(22);
\end{tikzpicture}
\caption{Schematic Galois correspondence for rings of modular forms. The diagram on the left represents the subgroup structure of $G$, the middle one the fields of meromorphic functions on modular curves generated by polynomials of the roots of the characteristic polynomial and the diagram on the right represents congruence subgroups that contain $\Gamma_{\mathrm{Galois}}$. The bottom line gives the explicit bijections between corresponding elements in the different diagrams. }
\label{tripleGalois}
\end{center}
\end{figure}
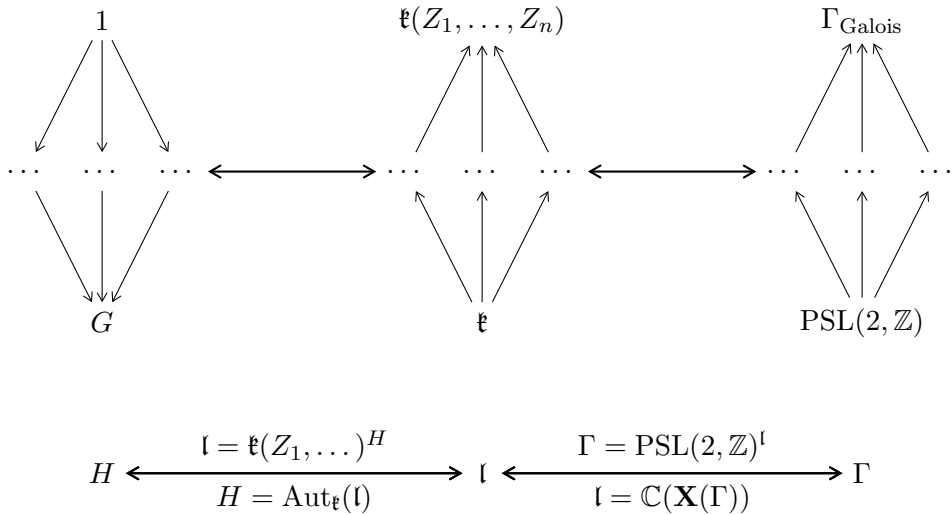
To the standard Galois duality that relates subgroups of the Galois group and field extensions, as depicted in the first two columns of figure \ref{tripleGalois}, we now add a third structure, namely the congruence subgroups containing $\Gamma_{\mathrm{Galois}}$. Let us then consider a congruence subgroup $\Gamma$ such that 
\begin{equation}
\label{inclusionGaloisCongruence}
   \Gamma_{\mathrm{Galois}} \subseteq \Gamma \subseteq \mathrm{PSL}(2,\mathbb{Z}) \, . 
\end{equation}
From $\Gamma$ we can construct a compact Riemann surface, called the \emph{modular curve} $\mathbf{X}(\Gamma)$, which is the appropriate compactification \cite{shimura,diamond} of $\Gamma \backslash \mathfrak{H}$. This permits us to use the general idea that an algebraic curve is entirely characterized by fields of functions $\mathbb{C}(\mathbf{X}(\Gamma))$ defined on it, and that field extensions correspond to morphisms between curves. It is clear that (\ref{inclusionGaloisCongruence}) implies 
\begin{equation}
    \mathbb{C}(\mathbf{X}(\Gamma_{\mathrm{Galois}})) \supseteq \mathbb{C}(\mathbf{X}(\Gamma)) \supseteq \mathbb{C}(\mathbf{X}(\mathrm{PSL}(2,\mathbb{Z}))) \, . 
\end{equation}
In other words, every group contained between $\Gamma_{\mathrm{Galois}}$ and $\mathrm{PSL}(2,\mathbb{Z})$ gives a field extension of $\mathbb{C}(\mathbf{X}(\mathrm{PSL}(2,\mathbb{Z})))$ contained inside $\mathbb{C}(\mathbf{X}(\Gamma_{\mathrm{Galois}}))$. Conversely, given such a field extension, the subgroup of $\mathrm{PSL}(2,\mathbb{Z})$ that leaves all its elements invariant contains $\Gamma_{\mathrm{Galois}}$, and in particular it is a congruence subgroup. 

To summarize, we have constructed a bijection between the subgroups of $G$, the field extensions between $\mathfrak{k}$ and $\mathfrak{k}(Z_1 ,\dots , Z_n)$, and the congruence subgroups that contain $\Gamma_{\mathrm{Galois}}$, as illustrated in figure \ref{tripleGalois}. In order to make the discussion more concrete, we work out an example in full detail.

\subsection{\texorpdfstring{The $\mathfrak{su}(2)$ ${\cal N}=1^\ast$ Theory}{}}
\label{SubsecSu2}

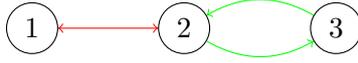
\begin{figure}
\begin{center}
\begin{tikzpicture}[node distance=3cm]
 \node (1) at  (0,0) [circle,draw] {1};
 \node (2) at  (1*2,0) [circle,draw] {2};
 \node (3) at  (2*2,0) [circle,draw] {3};
 \draw [->,green] (2) edge [bend right] (3);
 \draw [<->,red] (1) -- (2);
\draw [->,green] (3) edge [bend right] (2);
\end{tikzpicture}
\caption{The duality group action on the three vacua of the $\mathfrak{su}(2)$ ${\cal N}=1^\ast$ theory. The action of $T$ on the complexified coupling is indicated in green, the action
of $S$ in red.}
\label{su2duality}
\end{center}
\end{figure}

The simplest illustration of the general formalism above is provided by the ${\cal N}=1^\ast$ theory with gauge algebra $\mathfrak{su}(2)$. It has three inequivalent vacua on $\mathbb{R}^4$ \cite{Donagi:1995cf,Dorey:1999sj}. The vacua can be described in terms of the extremal values of an elliptic integrable potential. The corresponding extremal values of the superpotential are\footnote{We choose the normalization of the superpotential extremal values (\ref{su2values}) such that the characteristic polynomial (\ref{Psu2}) has coefficients in $\mathbb{Z}[E_4,E_6]$.} \cite{Dorey:1999sj,Polchinski:2000uf}
\begin{eqnarray}
W_1 &=& -2 E_2(q) +4 E_2(q^2) \, , \nonumber \\
W_2 &=& -2 E_2(q) + E_2 (q^{1/2})\, , \\
W_3 &=& -2 E_2(q) + E_2 (-q^{1/2})\,  , \nonumber
\label{su2values}
\end{eqnarray}
where the parameter $q=e^{2 \pi i \tau}$ is the exponential of the complexified coupling $\tau$ of the theory. The first value $W_{1}$ corresponds semi-classically to a Higgs vacuum obtained by giving a vacuum expectation value to the adjoint scalars while the two values $W_{2,3}$ correspond to confining vacua of an unbroken $\mathfrak{su}(2)$ pure ${\cal N}=1$ theory. We then compute the characteristic polynomial:
\begin{equation}
P_{\tau}^{\mathfrak{su}(2)}(W) = (W-W_1)(W-W_2)(W-W_3)
= W^3 - 3 E_4(q) W -2 E_6(q) \, . 
\label{Psu2}
\end{equation}
The discriminant of this degree three polynomial is 
\begin{equation}
\label{discriminantsu2}
   D (\tau)= 108 (E_4^3-E_6^2) \propto \Delta  (\tau) \, , 
\end{equation}
where $\Delta (\tau) = \frac{1}{1728}(E_4^3 - E_6^2 )$ is the modular discriminant, which has no zero on $\mathfrak{H}$. We are then in the situation where $D(\tau) $ has no non-trivial zero, and we can use the tools of section \ref{sectionTripleGalois}. Using the well-known transformation of the Eisenstein series $E_2$ under $\tau \mapsto -1/ \tau$, or physical expectations combined with consistency, one can derive the duality diagram of figure \ref{su2duality}, on which we read $c_2 = 1$, $c_3 = 0$, $\rho = 2+1$, $c_{\infty}=2$, and then $g=0$. As explained in footnote \ref{footnoteAssumption}, we assume that $\Gamma_{\mathrm{roots}}$ is a congruence subgroup, which allows us to identify it as
\begin{equation}
   \Gamma_{\mathrm{roots}} = \Gamma_0 (2) , 
\end{equation}
and then using the set of generators $\Gamma_0(2) = \langle T , S T^2 S \rangle$, one checks that it leaves $W_1$ invariant. The other roots $W_2$ and $W_3$ are left invariant under the two conjugates of $\Gamma_0(2)$, as can be read from the diagram: $W_2$ is left invariant by $S \Gamma_0 (2) S$ and $W_3$ by $(ST)^{-1} \Gamma_0 (2) (ST)$. One can check that this provides a set of coset representatives $ \{\alpha_1 = 1 , \alpha_2 = S, \alpha_3 = ST \}$ of $\Gamma_0 (2) \backslash  \mathrm{SL}(2,\mathbb{Z})$, 
\begin{equation}
\label{cosetRepGamma02}
 \mathrm{SL}(2,\mathbb{Z}) = \bigcup\limits_{i=1}^3 \Gamma_0 (2) \alpha_i  \, . 
\end{equation}
The three isomorphic stabilizers $\Gamma_0 (2)_i = \alpha_i^{-1 } \Gamma_0 (2) \alpha_i$ are the groups of matrices 
\begin{equation}
\label{genericMatricSL2}
   \left(\begin{array}{cc} a & b \\ c & d\end{array}  \right) \in \mathrm{SL}(2,\mathbb{Z})
\end{equation}
that satisfy, respectively, $c \equiv 0$, $b \equiv 0$ and $a-b+c-d \equiv 0$ modulo $2$. Using these characterizations, it is straightforward to prove that\footnote{A matrix in the intersection has $b \equiv c \equiv a-d \equiv 0$, and $ad \equiv 1$, hence $a \equiv d \equiv 1$ and it is in $\Gamma (2)$. The converse is obvious. } 
\begin{equation}
\label{Gamma2Intersection}
   \bigcap\limits_{i=1}^3 \Gamma_0 (2)_i = \Gamma (2) \, ,  
\end{equation}
which gives an alternative proof of the fact that $\Gamma (2) = \langle T^2 , S T^2 S \rangle$ is the stabilizer of all three superpotentials \cite{Ritz:2006ji}. The space of modular forms of weight two for $\Gamma (2)$ is two dimensional, and a basis is given by any family of two independent elements constructed from the $W_{1,2,3}$, for instance $(W_1, W_2-W_3)$. In fact, these two elements are also generators of the full ring of modular forms $\mathcal{M}(\Gamma(2))$.\footnote{One way to prove this is the following. According to example 1.7 in \cite{landesman2015spin}, the full field of modular forms is generated by the modular forms of weight up to $6$. The space $\mathcal{M}_4(\Gamma(2))$ of modular forms of weight four has dimension three and $\mathcal{M}_6(\Gamma(2))$ has dimension four, and we can check that these two spaces are generated as rings by $W_1$ and $W_2-W_3$.} 

\begin{figure}
\begin{center}
\begin{tikzpicture}[node distance=2cm]
\node (1) [emptybox] {$\mathfrak{S}_3$};
\node (2)[emptybox, above of=1, xshift=-1.5cm] {$ \langle (123) \rangle $};
\node (3)[emptybox, above of=2, xshift=+2cm] {$ \langle (13) \rangle $};
\node (4)[emptybox, right of=3, xshift=+1cm] {$ \langle (12) \rangle $};
\node (5)[emptybox, right of=4, xshift=+1cm] {$ \langle (23) \rangle $};
\node (6)[emptybox, above of=3] {$1$};
\draw [arrow] (2) -- node[anchor=east] {2}(1);
\draw [arrow] (3) -- node[anchor=east] {3}(1);
\draw [arrow] (4) -- node[anchor=east] {3 \, }(1);
\draw [arrow] (5) -- node[anchor=east] {3 \,   }(1);
\draw [arrow] (6) -- node[anchor=east] {3}(2);
\draw [arrow] (6) -- node[anchor=east] {2}(3);
\draw [arrow] (6) -- node[anchor=east] {2 \, }(4);
\draw [arrow] (6) -- node[anchor=east] {2 \, \, }(5);
\end{tikzpicture}
\caption{Subgroup structure of $\mathfrak{S}_3$. Each node is the subgroup that leaves invariant the corresponding field in figure \ref{fieldextensionssu2}. The integer on each arrow indicates the ratio of the cardinalities of the two groups connected by the arrow. }
\label{subgroupsS3}
\end{center}
\end{figure}
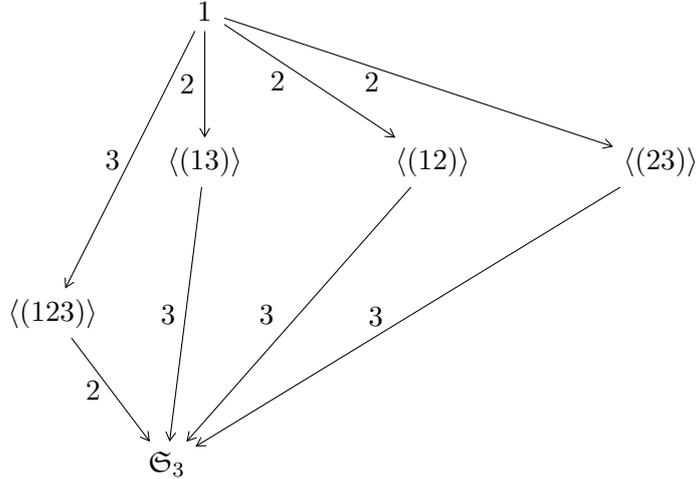

We now study the characteristic polynomial from the perspective of Galois theory. As by the general considerations of section \ref{generalitiesPolynomials}, the polynomial (\ref{Psu2}) has coefficients in the graded ring $\mathfrak{m}$ of modular forms. The polynomial $\tilde{P}_{\tau}^{\mathfrak{su}(2)}$, defined in equation (\ref{characteristicPolyWeight0}), is 
\begin{equation}
   \tilde{P}_{\tau}^{\mathfrak{su}(2)} = Z^3 - \frac{3j}{j-1728} Z - \frac{2j}{j-1728} \in \mathfrak{k} [Z] \, . 
\end{equation}
This polynomial is irreducible, and this gives degree three field extensions $\mathfrak{k} [Z_i]$ for $i=1,2,3$. Adding another root to one of the fields $\mathfrak{k} [Z_i]$, one obtains immediately the full extension $\mathfrak{k} [Z_1,Z_2,Z_3]$ because $Z_1 + Z_2 + Z_3 = 0$, and this extension has degree lower or equal to two as a consequence of $Z_1 Z_2 + Z_2 Z_3 + Z_3 Z_1 = -\frac{3j}{j-1728}$. But the degree can not be one, since $Z_2$ has fractional powers of $q$ in its expansion, and hence is not in $\mathfrak{k} [Z_1]$. So the degree is exactly two. 
\begin{figure}
\begin{center}
\begin{tikzpicture}[node distance=2cm]
\node (1) [emptybox] {$\mathfrak{k}$};
\node (2)[emptybox, above of=1, xshift=-1.5cm] {$\mathfrak{k}(\sqrt{j-1728})$};
\node (3)[emptybox, above of=2, xshift=+2cm] {$\mathfrak{k}(Z_2) $};
\node (4)[emptybox, right of=3, xshift=+1cm] {$\mathfrak{k}(Z_3)$};
\node (5)[emptybox, right of=4, xshift=+1cm] {$\mathfrak{k}(Z_1)$};
\node (6)[emptybox, above of=3] {$\mathfrak{k}(Z_1,Z_2,Z_3)$};
\draw [arrow] (1) -- node[anchor=east] {2}(2);
\draw [arrow] (1) -- node[anchor=east] {3}(3);
\draw [arrow] (1) -- node[anchor=east] {3 \, }(4);
\draw [arrow] (1) -- node[anchor=east] {3 \,   }(5);
\draw [arrow] (2) -- node[anchor=east] {3}(6);
\draw [arrow] (3) -- node[anchor=east] {2}(6);
\draw [arrow] (4) -- node[anchor=east] {2 \, }(6);
\draw [arrow] (5) -- node[anchor=east] {2 \, \, }(6);
\end{tikzpicture}
\caption{Field extensions generated by the roots of the characteristic polynomial for massive vacua of $\mathcal{N}=1^\ast$ theory with $\mathfrak{su}(2)$ gauge algebra. The integer on each arrow indicates the degree of the field extension. }
\label{fieldextensionssu2}
\end{center}
\end{figure}
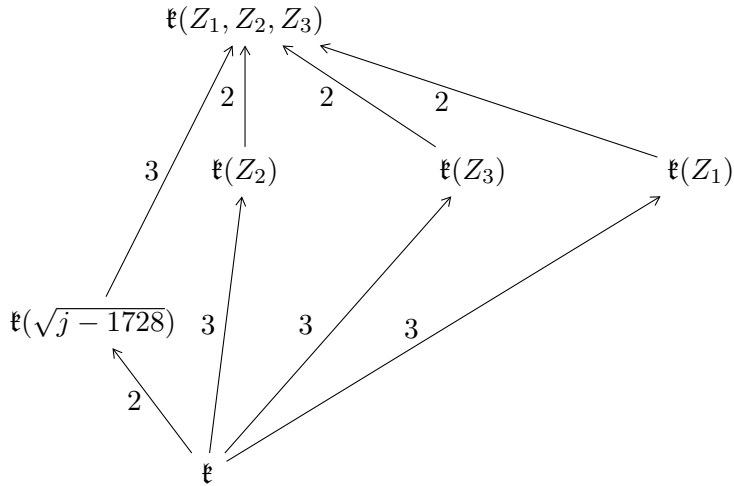

\begin{figure}
\begin{center}
\begin{tikzpicture}[node distance=2cm]
\node (1) [emptybox] {$\mathrm{PSL}(2,\mathbb{Z})$};
\node (2)[emptybox, above of=1, xshift=-1.5cm] {$\Gamma^2$};
\node (3)[emptybox, above of=2, xshift=+2cm] {$\Gamma_0(2)_2$};
\node (4)[emptybox, right of=3, xshift=+1cm] {$ \Gamma_0(2)_3$};
\node (5)[emptybox, right of=4, xshift=+1cm] {$\Gamma_0(2)$};
\node (6)[emptybox, above of=3] {$\Gamma(2)$};
\draw [arrow] (1) -- node[anchor=east] {2}(2);
\draw [arrow] (1) -- node[anchor=east] {3}(3);
\draw [arrow] (1) -- node[anchor=east] {3 \, }(4);
\draw [arrow] (1) -- node[anchor=east] {3 \,   }(5);
\draw [arrow] (2) -- node[anchor=east] {3}(6);
\draw [arrow] (3) -- node[anchor=east] {2}(6);
\draw [arrow] (4) -- node[anchor=east] {2 \, }(6);
\draw [arrow] (5) -- node[anchor=east] {2 \, \, }(6);
\end{tikzpicture}
\caption{Congruence subgroups between $\Gamma(1)$ and $\Gamma(2)$. The integer on each arrow indicate the degree the group at the end of the arrow inside the group at the origin.}
\label{CongruenceSubgroupsDiag}
\end{center}
\end{figure}
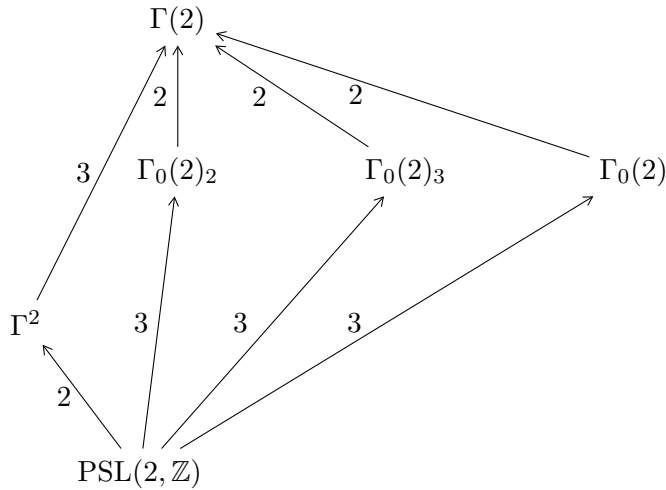

As a result of the Galois correspondence, there is a bijection between the diagram of ring extensions by polynomials of the roots of (\ref{Psu2}) and the subgroups of the Galois group $G^{\mathfrak{su}(2)}$, which is itself a subgroup of the group of permutations of these roots. The computations of the previous paragraph show that $|G^{\mathfrak{su}(2)}| \geq 2 \cdot 3$, which implies 
\begin{equation}
   G^{\mathfrak{su}(2)} = \mathfrak{S}_3 \, . 
\end{equation}
We then obtain immediately the diagram of subgroups, see figure \ref{subgroupsS3}. In this first example then, the Galois group is equal to the full permutation group of the roots. Moreover, the Galois group coincides with the duality group action on the massive vacua. According to the Galois correspondence, we observe that there should also be a degree two extension of $\mathfrak{k}$ given by the functions that are invariant under the permutation $(123)$, but not under the exchange of only two roots. An example of a function leading to such an extension is 
\begin{equation}
    (Z_1-Z_2)(Z_2-Z_3)(Z_3-Z_1) \propto \frac{j}{(j-1728)^{3/2}} \, , 
\end{equation}
which indeed is of order two. The function $\sqrt{j-1728}$ is sent to $-\sqrt{j-1728}$ by the transformations $S$ and $T$, so its stabilizer is the group $\Gamma^2 = \langle T^2 , ST \rangle$ of elements of $\mathrm{PSL}(2,\mathbb{Z})$ of even length, when considered as words with letters $S,T$. This allows to complete the diagram of field extensions, see figure \ref{fieldextensionssu2}. In turn, we have computed the stabilizer group of all these field extensions, which gives birth to a third diagram where the vertexes are certain congruence subgroups, as represented in figure \ref{CongruenceSubgroupsDiag}. This completes the explicit Galois correspondence of figure \ref{tripleGalois} in the $\mathfrak{su}(2)$ case.

\subsection{Applications}

In this subsection, we pause to consider a few applications of the constructions we presented.
In section \ref{SectionSun}, we provide further examples.

\subsubsection{Tensor products}

We have introduced a characteristic polynomial 
of degree $n$ whose roots are the values of $W$
in the $n$ vacua of the theory $\mathcal{T}$. This polynomial can be interpreted as
living in the tensor product of theories $\mathcal{T}^{\otimes n}$. 
We are thus lead to explore the tensor products of the theory $\mathcal{T}$ with itself. Some special diagonal sectors of this extended theory (with a different
choice of vacuum for each factor) have a reduced set of vacuum expectation values for the superpotential, that correspond to non-trivial subgroups of the Galois group. In other words, the Galois correspondence tells us what are the sectors within the tensor product that have an enhanced modular invariance. This provides an incarnation of the Galois correspondence in tensor product theories with a particular choice of boundary conditions in each factor theory.

\subsubsection{On the Chiral Ring}

In this subsection, we consider a few properties of the chiral ring. Firstly, we take the perspective that we give two of the three adjoint chiral multiplets a larger mass than the third. After having integrated out those two chiral multiplets, the effective superpotential can be expressed in term of the third one, which we call $\Phi$, as
\begin{equation}
\label{operatorW}
W = \langle \mathrm{tr} \,  \Phi^2 \rangle \, . 
\end{equation}
The chiral ring is then generated by the operators  $\mathrm{tr} \, \Phi^k$, $\mathrm{tr} \, \Phi^k W_\alpha $ and $\mathrm{tr} \, \Phi^k W_\alpha W^\alpha$ where $W_\alpha$ is the vector multiplet superfield. We focus on the bosonic sector. 
In the $\mathfrak{su}(N)$ theory, there are $N$ operators $\mathrm{tr} \, \Phi^k$ and $N$ operators $\mathrm{tr} \, \Phi^k W_\alpha W^\alpha$ which satisfy quantum chiral ring relations. 
Let us study the vacuum expectation values of $\mathrm{tr} \, \Phi^k$ first.

There are several possible ways to define condensates $\langle \mathrm{tr} \,  \Phi ^k \rangle$ in the $\mathcal{N}=1^\ast$ theory. One of them is to use the Dijkgraaf-Vafa approach and solve a matrix model \cite{Dijkgraaf:2002fc} and another is to use the exact integrable system and compute traces of powers of the Lax matrices \cite{Dorey:2002ad}. However, mixing ambiguities \cite{Aharony:2000nt} will makes these different results for $\langle \mathrm{tr} \, \Phi ^k \rangle$ differ by linear combinations in the operators $\langle \mathrm{tr} \, \, \Phi ^l \rangle$ for $l<k$ as well as vacuum independent functions of the coupling $\tau$. This ambiguity can be solved when there are no non-trivial zeros by choosing the  combination that behaves as a modular form under the modular transformations \cite{Dorey:2002tj}. 
In this case, this proves that the sector of the chiral ring generated by the $\langle \mathrm{tr} \, \, \Phi ^k \rangle$ is included in the space of forms $\mathcal{M}(\Gamma_{\textrm{Galois}})$. 

The full chiral ring also contains quasi-modular forms, as for instance $\langle \mathrm{tr} \,  W^{\alpha} W_{\alpha} \rangle = -\frac{1}{2 \pi i} \partial_{\tau} \langle \mathrm{tr} \, \Phi ^2 \rangle $. 
As we know the value of the exact effective superpotential $\langle W \rangle$ in a vacuum, we can deduce the value of the gaugino condensate because it appears in the tree-level superpotential multiplied by $\tau$. We then have $\langle S \rangle=\langle \mathrm{tr} \,  W^{\alpha} W_{\alpha} \rangle = -\frac{1}{2 \pi i} \partial_{\tau} \langle W \rangle $.
But we also know from our general discussion that there is a polynomial equation $P( \langle W \rangle , E_4   , E_6  )=0$, where $P$ is a polynomial with integer coefficients. We can differentiate with respect to $\tau$: 
\begin{equation}
-  \langle S \rangle \partial_{\langle W \rangle} P + \frac{1}{3} (E_2 E_4 - E_6) \partial_{E_4} P + \frac{1}{2} (E_2 E_6 - E_4^2)  \partial_{E_6} P = 0 \, . 
\end{equation}
This equation shows that $\langle S \rangle$ belongs to the field $\mathfrak{k}(E_2,W_i) = \mathbb{C}(E_2,E_4,E_6,W_i)$: 
\begin{equation}
\label{vevS}
  \langle S \rangle = \frac{\frac{1}{3} (E_2 E_4 - E_6) \partial_{E_4} P + \frac{1}{2} (E_2 E_6 - E_4^2)  \partial_{E_6} P }{ \partial_{\langle W \rangle} P}\, . 
\end{equation}
In fact, we can say more: the derivative of any modular form in $\mathcal{M}(\Gamma)$ belongs to the ring of quasi-modular forms $\mathcal{M}^{\mathrm{quasi}}(\Gamma)$ that is obtained by simply adding $E_2$, 
\begin{equation}
\mathcal{M}^{\mathrm{quasi}}(\Gamma) = \mathcal{M}(\Gamma)[E_2] \, . 
\end{equation}
As a consequence, although it is not obvious on the expression (\ref{vevS}), the gaugino condensate can be expressed as a linear combination of $E_2^2$, $E_2 W$ and $W^2$, and the explicit combination is easily deduced in a given example from (\ref{vevS}). For instance, in the $\mathfrak{su}(2)$ case, the polynomial (\ref{Psu2}) and the division in the corresponding ideal give
\begin{equation}
    \langle S \rangle = - \frac{(E_2 E_4 - E_6) W + E_2 E_6 - E_4^2 }{3 (W^2-E_4)} = \frac{1}{6} (W^2 - W E_2 - 2 E_4) \, . 
\end{equation}
It would be most interesting to complete the calculation of the vacuum expectation values of the chiral ring operators, including the vacuum expectation values of all three chiral fields $\Phi_i$.


\subsubsection{Domain Wall Kinematics in the Quantum Theory} 
\label{domainwallkinematics}
There are also very direct uses of the superpotential values.
To illustrate this aspect, we analyze the kinematical
stability of BPS domain walls of the $\mathfrak{su}(2)$ ${\cal N}=1^\ast$ theory in the quantum theory,
at any value of the complexified gauge coupling $\tau$. We will
combine our results with those obtained in the classical theory \cite{Bachas:2000dx},
and the physics of confining gauge theories to propose a global picture of the existence of domain walls in the theory.

Firstly, we recall that half BPS domain walls in ${\cal N}=1$ theories in four dimensions preserve two supercharges.
The corresponding central charge is a complexified tension $T$, which relates to the tension $M$ through the formula \cite{Abraham:1990nz}
\begin{eqnarray}
M &\ge& |T| \, ,
\end{eqnarray}
with saturation for BPS domain walls.
If a BPS domain wall composed of BPS domain walls with complexified tensions $T_{1}$ and $T_2$ exists, it then has mass
\begin{eqnarray}
M_3 &=& |T_3|=|T_1+T_2| \le |T_1|+|T_2| =M_1+M_2
\, ,
\label{kinstab}
\end{eqnarray}
which is automatically stable. The wall is marginally stable if the phases of the central charges are equal. This will typically happen in a space of real co-dimension one in the space
of couplings.
A generic expectation would be then that BPS domain walls are stable or unstable on one or another side of such a wall of marginal stability \cite{Abraham:1990nz,Ritz:2003qq}.

Let's reconsider the example of the $\mathfrak{su}(2)$ theory with superpotential
values $W_{1,2,3}$ as in equation (\ref{su2values}).
We note that for $\tau$ purely imaginary, as well as in the semi-classical
limit $\tau \rightarrow i \infty$, the critical values are real.
We have available an existence analysis for classical domain walls \cite{Bachas:2000dx}.
In this limit, one has a full handle on the kinetic term, and therefore on the 
first order flow equation for a BPS domain wall \cite{Abraham:1990nz}. 
One can then explicitly determine whether the BPS wall exists.
In \cite{Bachas:2000dx} it was found that
a domain wall exists between two massive vacua
if the superpotential on the left (say) is larger than the one on the right.
Thus, for $\mathfrak{su}(2)$ for instance, we have a supersymmetric wall interpolating between the Higgs vacuum and the confining vacua. Indeed, the classical limits of the above superpotentials are
$(W_1,W_2,W_3)_{\mbox{\tiny clas}}=(-1,-1,2)$ respectively. 
The first two correspond to confining vacua and the last to the Higgs vacuum. The first two are degenerate in the classical regime. Thus, the classical limit suggests the existence of supersymmetric domain walls between vacua $1$ and $3$ and between $2$ and $3$. 
In the classical limit, the tension of the domain wall between the (quantum mechanically)
confining vacua becomes zero. At finite and small coupling, the expectation from
the pure ${\cal N}=1$ theory is that the domain wall between the two confining vacua
exists and is BPS. 
At finite and small coupling,  
these facts provide us with a full picture of the domain walls. We can ask 
whether this picture is accurate in the deep quantum regime.

Firstly, we check whether we have another point at which extremal values of the superpotential coincide (as they do in the classical point). According to (\ref{discriminantsu2}), there are coinciding roots at values of $\tau$ where $D(\tau)$ vanishes, which happens only at the cusps.  

Most importantly, we can also scan for walls of marginal stability, where the three roots of the 
characteristic polynomial are aligned. These may signal that BPS walls that exist near the classical regime decay at strong coupling. The sum  of the roots
of the characteristic polynomial always vanishes. If they are aligned, then there exists a phase $\theta$ such that the $e^{-i \theta} W_j$ are all real. Those are then the three real roots of  $W^3 - 3 e^{2 i \theta} E_4 W - 2 e^{3 i \theta} E_6 $. Hence the coefficients of this polynomial must be real, and the discriminant must be positive. In particular $(e^{2 i \theta} E_4)^3 / (e^{3 i \theta} E_6)^2 = E_4^3/E_6^2$ is real. This is true on the lines $\tau_1 = 0$ and $\tau_1 = 1/2$. However on the line $\tau_1 = 1/2$, the discriminant $\eta(\tau)^{24}$ is negative. Therefore the three roots are aligned if and only if $\tau_1 = 0$ (in the fundamental domain). This potentially is a wall of marginal stability. 
In any event, since the classical regime near $\tau_2 = i \infty$, at non-zero $\theta$ angle extends towards strong coupling without crossing the only potential wall of marginal stability, we conclude that the existence of BPS domain walls in the deep quantum regime is  fixed by the existence analysis of BPS domain walls in the classical regime at non-zero $\theta$-angle. This conclusion illustrates that 
using our knowledge of the exact values of the superpotential, we can infer properties of the deep quantum regime of the ${\cal N}=1^\ast$ theory.

\section{\texorpdfstring{The $\mathfrak{su}(N)$ Theory}{}}
\label{SectionSun}

In this section, we generalize to the $\mathfrak{su}(N)$ theory the structures covered in the previous section in the case of $\mathfrak{su}(2)$. We will see that the Galois group is in general  smaller than the full permutation group of vacua. The order $N$ lattices of $\mathbb{Z}_N \times \mathbb{Z}_N$ play a classifying role \cite{Donagi:1995cf}, and we start by collecting useful properties of these lattices. These properties imply that all $\mathfrak{su}(N)$ $\mathcal{N}=1^\ast$ theories belong to the class of theories that still display reduced modular invariance in any given vacuum. See also \cite{Aharony:2000nt}. This can be thought
off as a consequence of the toroidal geometries governing
both the ${\cal N}=2$ and the ${\cal N}=1$ mass deformed
theories \cite{Donagi:1995cf}. We detail the reduced modular invariance in low-rank examples and highlight some salient features. 

\subsection{\texorpdfstring{Order $N$ Lattices of $\mathbb{Z}_N \times \mathbb{Z}_N$ and Dualities}{}}
\label{subsecSubgroupsZN2}

An order $N$ lattice of the group $\mathbb{Z}_N \times \mathbb{Z}_N$ is 
an order $N$ subgroup of $\mathbb{Z}_N \times \mathbb{Z}_N$. There are $\sigma_1(N)=\sum_{d|N} d$ such lattices. We use the following parametrization: 
\begin{equation}
\label{equationLattice}
L = L_{N,p,k} = \left\{ (pr+sk , sq) \in \mathbb{Z}_N^2 | r \in \mathbb{Z}_q , \,   s \in \mathbb{Z}_p \right\} \, , 
\end{equation}
where $p$ is a divisor of $N$, the integer $q$ is defined by $pq=N$ and $k \in \mathbb{Z}_p$. Moreover, we define the \emph{level} of the lattice $L_{N,p,k}$ to be $d = \textrm{gcd} (p,k,N/p)$. Note that we always have $d^2 | N$. The group $\mathrm{SL}(2,\mathbb{Z})$ acts on the set of lattices by left multiplication on the lattice points. 

The motivation to study the modular transformation properties of these lattices is twofold. On the one hand, the lattices label the possible choices of line operators in a given theory \cite{Aharony:2013hda}, and hence label the theories with $\mathfrak{su}(N)$ gauge algebra. More details on this aspect can be found in appendix \ref{appendixTheoriessuN}. 

More important to us here is the fact that the order $N$ lattices of $\mathbb{Z}_N \times \mathbb{Z}_N$ are in bijection with the massive vacua of the $\mathcal{N}=1^\ast$ $\mathfrak{su}(N)$ theory on $\mathbb{R}^4$ \cite{Donagi:1995cf}. The superpotential in the vacuum corresponding to the lattice $L$ is given by \cite{Dorey:1999sj,Polchinski:2000uf}
\begin{equation}
   W_L = \sum_{A \neq B \in L} \wp \left(\frac{A-B}{N} , \tau \right) \, , 
\end{equation}
where we identify a point $A = (x,y) \in  L$ with the complex number $A = x + \tau y$. Using a re-parametrization of the sum, one obtains 
\begin{equation}
  Z_L = \frac{E_4}{E_6} W_L = N  \frac{E_4}{E_6} \sum_{A \in L- \{(0,0)\}} \wp \left(\frac{A}{N} , \tau \right)   \, .  
\end{equation}
Two questions are particularly relevant for our study of the permutation of the massive vacua: 
\begin{itemize}
    \item Which lattices can be related by a duality transformation?
    \item What are the duality transformations that leave a given lattice (respectively, the set of all lattices) invariant?
\end{itemize}
The answer to the above questions is given by the three following properties of the lattices (\ref{equationLattice}). 
\begin{enumerate}
    \item The structure of $T$-multiplets is as follows: each divisor $D$ of $N$ gives $\mathrm{gcd}(D,\frac{N}{D})$ cycles of size $D/\mathrm{gcd}(D,\frac{N}{D})$. 
    \item Two lattices are related by a duality if and only if they have the same level. Hence the number of duality multiplets is the number of divisors of $N$ that are squares. 
    \item Any lattice of level $d$ is left invariant by $\alpha^{-1} \Gamma_0(N/d^2) \alpha $, where $\alpha$ is a coset representative of $\Gamma_0(N/d^2)$ in $\mathrm{SL}(2,\mathbb{Z})$, and this gives a bijection between level $d$ lattices and $\mathrm{SL}(2,\mathbb{Z})/\Gamma_0(N/d^2)$. 
\end{enumerate}
\label{propertiessection}
The proofs of these properties are relegated to appendix \ref{AppendixProofs}. 
Note that it is clear that $\Gamma(N)$ leaves all the lattices invariant (indeed, any matrix in $\Gamma(N)$ acts as the identity in $\mathbb{Z}_N^2$ by definition).

 We give one example. For $N=4$, property 1 tells us that we have three cycles of size one and one cycle of size four under $T$. As a result of property 2, these cycles organize themselves into two multiplets, which by property 3 have respectively size one and six.

\subsection{The Characteristic Polynomial and the Galois Group}

The structure of the characteristic polynomial is deduced from the structure of the duality diagram of the vacua. According to property 2 in subsection \ref{subsecSubgroupsZN2}, the number of connected components of this graph of dualities is exactly equal to the number of squares dividing $N$, and from property 3 the number of vertexes in each component is equal to the index of $\Gamma_0(N/d^2)$. This gives the following decomposition of the number of vacua
\footnote{A dual way of saying the same thing is 
\begin{equation}
\sum\limits_{P' \subset P(N)} (-1)^{|P'|} \sigma_1 \left( N \prod\limits_{p \in P'} p^{-2} \right) = N \prod\limits_{p|N} \left( 1+ \frac{1}{p}\right) \, ,
\end{equation}
where $P(N)$ is the set of primes $p$ whose square divides $N$. }
\begin{equation}
\label{decompositionMultiplets}
 \sigma_1 (N)  = \sum\limits_{d^2 | N} \frac{N}{d^2} \prod\limits_{p | \frac{N}{d^2}} \left(1+\frac{1}{p} \right) \, ,  
\end{equation}
and this decomposition is reflected into the factorization of the degree $ \sigma_1 (N) $ characteristic polynomial (\ref{characteristicPoly}) into factors of degrees $\frac{N}{d^2} \prod_{p | \frac{N}{d^2}} \left(1+\frac{1}{p} \right)$. Each multiplet is characterized by its own permutation group, which is the Galois group of the associated factor in the polynomial. This is again reminiscent of the characterization of phases of ${\cal N}=1$ theories through the Galois group associated to the chiral ring relations, advocated in \cite{Ferrari:2008rz}. 
In general the full Galois group is a subgroup of the product of the individual Galois groups. 
The physics in these multiplets (e.g. the spectrum of massive excitations) will be different, while in a given multiplet, it will be the same by duality.\footnote{Similar remarks apply to the space of theories discussed in appendix \ref{appendixTheoriessuN}.}

The identification between the roots of the polynomial and the order $N$ subgroups of $\mathbb{Z}_N \times \mathbb{Z}_N$, on which there is a well-defined action of the modular group, indicates  that $\sqrt{D}$ has no branch cut. Accordingly, we know  that the field extension is a subfield of the field of modular functions of the modular curve of $\Gamma(N)$ \cite{Ritz:2006ji}. In the next paragraph, we examine in detail the relation between these two fields, and use the Galois correspondence to learn about the structure of the Galois group.

To compute the Galois group, one follows the same steps as for the $\mathfrak{su}(2)$ example in section \ref{SubsecSu2}. We compute the intersection of the stabilizers of all the $W_i$. Let $(\alpha_j)$ be a family of coset representatives of $\Gamma_0(N)$ in $\mathrm{SL}(2,\mathbb{Z})$. We clearly have the inclusion inspired from (\ref{Gamma2Intersection})
\begin{equation}
\label{intersectionGammaN}
   \bigcap\limits_{j} \alpha_j^{-1} \Gamma_0(N) \alpha_j \supseteq \Gamma (N) \, . 
\end{equation}
The inclusion can be strict. To see this, we first prove that if
$\alpha_i$ equals the matrix
\begin{equation}
   \alpha_i = \left(
\begin{array}{cc}
 e & f \\
 g & h \\
\end{array}
\right) \, , 
\end{equation}
then we have the coset
\begin{equation}
\label{conjugateGamma0N}
    \alpha_j^{-1} \Gamma_0(N) \alpha_j = 
    \left\{  \left(\begin{array}{cc} a & b \\ c & d\end{array}  \right) \in \mathrm{SL}(2,\mathbb{Z}) |  g^2 b +gh (d-a) -h^2 c \equiv 0  \textrm{ mod } N \right\}  \, . 
\end{equation}
If $N \geq 2$, we always have at least the three coset representatives $1,S,ST^{-1}$ in the list $(\alpha_j)$, as can be read on the duality diagram. This shows that the left-hand side of (\ref{intersectionGammaN}) is included in 
\begin{equation}
    \Gamma_{\textrm{Galois}} (N)  := \left\{  \left(\begin{array}{cc} a & b \\ c & d\end{array}  \right) \in \mathrm{SL}(2,\mathbb{Z}) | b \equiv c \equiv a-d \equiv a^2 -1 \equiv 0  \textrm{ mod } N \right\} \, ,
\end{equation}
but the general expression (\ref{conjugateGamma0N}) shows that this group is a subgroup of all the $\alpha_j^{-1} \Gamma_0(N) \alpha_j$. Therefore we have proven that 
\begin{equation}
\label{GaloisCongGroupsuN}
   \bigcap\limits_{j} \alpha_j^{-1} \Gamma_0(N) \alpha_j  = \Gamma_{\textrm{Galois}} (N) \supseteq \Gamma (N) \, , 
\end{equation}
and the index of the inclusion is the number $s(N)$ of solutions of the equation $a^2 -1 \equiv 0  \textrm{ mod } N$.\footnote{If $\omega(N)$ is the number of distinct prime divisors of $N$, we have 
\begin{equation}
    s(N) = \begin{cases} 
    2^{\omega(N) -1} & \textrm{ for } N \equiv \pm 2 \textrm{ mod } 8 \\
    2^{\omega(N) } & \textrm{ for } N \equiv \pm 1 , \pm 3 , 4 \textrm{ mod } 8 \\
    2^{\omega(N) +1} & \textrm{ for } N \equiv 0 \textrm{ mod } 8 \, . 
    \end{cases}
\end{equation}} 
Using the index of $\Gamma (N)$ in $\mathrm{SL}(2,\mathbb{Z})$ (see figure \ref{congSub} in the appendix) and the Galois correspondence, we deduce that 
\begin{equation}
\label{cardinalityGaloisGroup}
    |G_{\mathfrak{su}(N)}| = [\mathrm{SL}(2,\mathbb{Z}) : \Gamma_{\textrm{Galois}} (N)] = \frac{N^2 \phi (N)}{s(N)} \prod\limits_{p|N} \left( 1+ \frac{1}{p}\right) \, . 
\end{equation}
This formula simplifies if $N \geq 3$ is prime: in this case, $\mathbb{Z}_N$ is a field where in addition $1 \neq -1$, so $s(N)=2$, and we obtain 
\begin{equation}
   |G_{\mathfrak{su}(N)}|_{N \geq 3 \textrm{ prime}} = \frac{N^3-N}{2} = [\mathrm{PSL}(2,\mathbb{Z}) : \Gamma (N)]_{N \textrm{ prime}}\, . 
\end{equation}

Although it is difficult to characterize precisely the Galois group in general, it is clear that it is a group much smaller than $\mathfrak{S}_{\#  \textrm{vacua}}$. For instance we have just seen that for the $\mathfrak{su}(N)$ theory with $N \geq 3$ prime, the ratio of orders of $\mathfrak{S}_{\# \textrm{vacua}}$ and $G$ is $2(N-2)!$. For a given value of $N$, the Galois group can be computed explicitly as the group generated by the two transformations $S$ and $T$ acting on the order $N$ subgroups of $\mathbb{Z}_N \times \mathbb{Z}_N$ as described in subsection \ref{subsecSubgroupsZN2}.

\subsection{\texorpdfstring{The $\mathfrak{su}(3)$ Example}{}}
\label{su3}

Let us apply the general results to the $\mathfrak{su}(3)$ example. In this case, the result (\ref{cardinalityGaloisGroup}) gives immediately the congruence subgroup $\Gamma_{\textrm{Galois}} = \Gamma(3)$, the cardinality of the Galois group, $|G_{\mathfrak{su}(3)}| = 12 = \frac{1}{2} |\mathfrak{S}_4|$, and therefore the Galois group itself 
is \begin{equation}
   G_{\mathfrak{su}(3)} = \mathfrak{A}_4 \, ,  
\end{equation}
the alternating group $\mathfrak{A}_4$ of even permutations. 

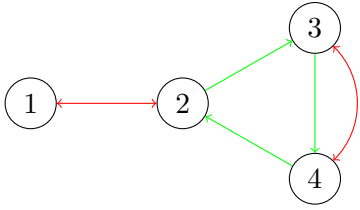
\begin{figure}
\begin{center}
\begin{tikzpicture}[node distance=3cm]
 \node (1) at  (0,0) [circle,draw] {1};
 \node (2) at  (1*2,0) [circle,draw] {2};
 \node (3) at  (1.87*2,.5*2) [circle,draw] {3};
 \node (4) at  (1.87*2,-.5*2) [circle,draw] {4};
 \draw [->,green] (2) -- (3);
 \draw [->,green] (3) -- (4);
 \draw [->,green] (4) -- (2);
 \draw [<->,red] (1) -- (2);
  \draw [<->,red] (3) to [out=-45,in=45] (4);
\end{tikzpicture}
\caption{The duality group action on the four vacua of the $\mathfrak{su}(3)$ ${\cal N}=1^\ast$ theory.}
\label{su3duality}
\end{center}
\end{figure}

This is confirmed by an explicit computation based on order $3$ sublattices of $\mathbb{Z}^2$, which gives the duality diagram of figure \ref{su3duality}, that can also be interpreted as the coset graph of $\Gamma_0(3)$. Physically these are four isolated extrema that correspond to one Higgs vacuum and three confining vacua. The permutations $T=(234)$ and $S=(12)(34)$ indeed generate $\mathfrak{A}_4$. 

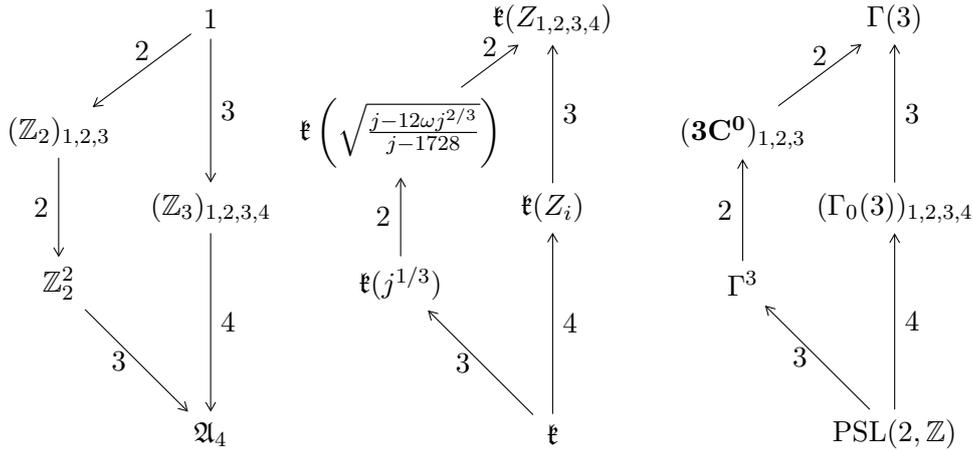
\begin{figure}
\begin{center}
\begin{tikzpicture}[node distance=2cm]
\node (1) [emptybox] at (0,0) {$\mathfrak{A}_4$};
\node (2)[emptybox, above of=1, xshift=-2cm] {$\mathbb{Z}^2_2$};
\node (3)[emptybox, above of=2, xshift=+2cm, yshift=-1cm] {$(\mathbb{Z}_3)_{1,2,3,4}$};
\node (5)[emptybox, above of=2 ] {$(\mathbb{Z}_2)_{1,2,3}$};
\node (10)[emptybox, above of=3,yshift=.5cm] {$1$};
\draw [arrow] (2) -- node[anchor=east] {3}(1);
\draw [arrow] (3) -- node[anchor=west] {4}(1);
\draw [arrow] (5) -- node[anchor=east] {2}(2);
\draw [arrow] (10) -- node[anchor=west] {3}(3);
\draw [arrow] (10) -- node[anchor=south] {2}(5);
\node (11) [emptybox] at (4.5,0) {$\mathfrak{k}$};
\node (12)[emptybox, above of=11, xshift=-2cm] {$ \mathfrak{k}(j^{1/3})$};
\node (13)[emptybox, above of=12, xshift=+2cm, yshift=-1cm] {$ \mathfrak{k}(Z_i)$};
\node (15)[emptybox, above of=12 ] {$\mathfrak{k} \left( \sqrt{\frac{j-12 \omega j^{2/3}}{j-1728}} \right)$};
\node (110)[emptybox, above of=13,yshift=.5cm] {$\mathfrak{k}(Z_{1,2,3,4})$};
\draw [arrowb] (12) -- node[anchor=east] {3}(11);
\draw [arrowb] (13) -- node[anchor=west] {4}(11);
\draw [arrowb] (15) -- node[anchor=east] {2}(12);
\draw [arrowb] (110) -- node[anchor=west] {3}(13);
\draw [arrowb] (110) -- node[anchor=south] {2}(15);
\node (21) [emptybox] at (9,0) {$\mathrm{PSL}(2,\mathbb{Z})$};
\node (22)[emptybox, above of=21, xshift=-2cm] {$\Gamma^3$};
\node (23)[emptybox, above of=22, xshift=+2cm, yshift=-1cm] {$(\Gamma_0(3))_{1,2,3,4}$};
\node (25)[emptybox, above of=22 ] {$(\mathbf{3C^0})_{1,2,3}$};
\node (210)[emptybox, above of=23,yshift=.5cm] {$\Gamma(3)$};
\draw [arrowb] (22) -- node[anchor=east] {3}(21);
\draw [arrowb] (23) -- node[anchor=west] {4}(21);
\draw [arrowb] (25) -- node[anchor=east] {2}(22);
\draw [arrowb] (210) -- node[anchor=west] {3}(23);
\draw [arrowb] (210) -- node[anchor=south] {2}(25);
\end{tikzpicture}
\caption{Summary of the Galois correspondence for gauge algebra $\mathfrak{su}(4)$. Here $\omega$ is a third root of unity.}
\label{subgroupsA4}
\end{center}
\end{figure}

The subgroup structure of $\mathfrak{A}_4$ is presented on the left of figure \ref{subgroupsA4}. The subgroups $\mathbb{Z}_2$ are generated by permutations of the form $(ab)(cd)$, and we have added a subscript $1,2,3$ to represent the three subgroups of this kind. Similarly, there are four subgroups isomorphic to $\mathbb{Z}_3$ corresponding to four independent three-cycles. Finally, we have the group $\mathbb{Z}_2^2 = \{1 , (12)(34) , (13)(24) , (14)(23)\}$. Using the extremal values of the superpotential, we can compute the  characteristic polynomial for the $\mathfrak{su}(3)$ quadruplet 
\begin{equation}
\label{polysu3}
P^{\mathfrak{su}(3)}(W) = W^4 - 8 E_6 W - 6 E_4 W^2 - 3 E_4^2 \, . 
\end{equation}
As a check, the discriminant 
\begin{equation}
\label{discriminantsu3}
   D =-2^{24}3^9 \Delta^2
\end{equation}
is the square of a modular form, which confirms that $\sqrt{D}$ has no branch cut.

The Galois correspondence relates the field extensions generated by polynomials of the roots $Z_i$ to congruence subgroups that contain $\Gamma_{\textrm{Galois}}(3) = \Gamma(3)$. Those are represented on the right of figure \ref{subgroupsA4}, using the notations of \cite{cummins}. The corresponding fields of modular functions are then engendered by polynomials of the roots left invariant by the associated subgroup of $\mathfrak{A}_4$.
For instance, $\mathbb{C}(\mathbf{X}(\Gamma^3))$ is generated by polynomials left invariant by $\mathbb{Z}_2^2$, like
\begin{equation}
   (Z_1 -Z_2) (Z_3 - Z_4) + (Z_1 - Z_3) (Z_2 - Z_4) + (Z_1 - Z_4) (Z_3 - Z_2) \sim  \frac{j^{2/3}}{j-1728} \, . 
\end{equation}
The extension of $ \mathfrak{k}$ generated by this function is $\mathfrak{k}(j^{1/3})$, and the fact that is has degree three guarantees that it is the right extension. 
Similarly, 
\begin{equation}
   Z_1 + Z_2 - Z_ 3 - Z_4 \sim \sqrt{\frac{j-12j^{2/3}}{j-1728}} \, 
\end{equation}
is associated to the group $\mathbb{Z}_2 = \{1 , (12)(34)\}$ and to one of the three congruence subgroups in the isomorphism class $\mathbf{3C^0}$.

\subsection{\texorpdfstring{The $\mathfrak{su}(4)$ Example}{}}

The $\mathfrak{su}(4)$ example is the first one where we have two disjoint multiplets, according to property 2 of section \ref{propertiessection}. The cardinality of the Galois group is given by (\ref{cardinalityGaloisGroup}), $ |G_{\mathfrak{su}(4)}| =  24= [\mathrm{PSL}(2,\mathbb{Z}) : \Gamma (4)]$, and we have $\Gamma_{\textrm{Galois}}(4) = \Gamma(4)$. 
In contrast with the situation encountered in the $\mathfrak{su}(3)$ example, the cardinality is not sufficient to determine the Galois group, because there are five conjugacy classes of cardinal $24$ groups in $\mathfrak{S}_6$. However, using our correspondence with congruence subgroups, we can extract more information about the subgroups of $G_{\mathfrak{su}(4)}$. In particular, using the facts collected in \cite{cummins}, one can compute the diagram of congruence subgroups between $\mathrm{PSL}(2,\mathbb{Z})$ and $\Gamma(4)$. There are 30 subgroups that can be organized into 11 classes up to conjugation. Moreover, we know from subsection \ref{subsecSubgroupsZN2} that all the six vacua are permuted by the Galois group. These properties uniquely identify the Galois group,
\begin{equation}
   G_{\mathfrak{su}(4)}   \cong \mathfrak{S}_4 \, . 
\end{equation}
This result can be confirmed using the known duality permutations on massive vacua. 

Let us show that the Galois correspondence leads to a complete understanding of the behavior of the vacua under dualities in figure \ref{fullDiagsu4}. The diagram in figure \ref{fullDiagsu4} represents the subgroup structure of the Galois group $\mathfrak{S}_4$, the congruence groups contained between $\Gamma_{\mathrm{Galois}} = \Gamma(4)/\{\pm 1\}$ and $\mathrm{PSL}(2,\mathbb{Z})$, and the associated function field extensions.

\begin{figure}
\begin{center}
\begin{tikzpicture}[x=2cm,y=1cm]
\node (0)[emptybox] at (0,-1) {};
\node (1)[emptybox] at (0,0) {$\mathrm{PSL}(2,\mathbb{Z})$};
\node (2)[emptybox] at (-2,4) {$\mathbf{4A^0}$};
\node (3)[emptybox] at (0,2) {$\Gamma^2$};
\node (4)[emptybox] at (2,3) {$\Gamma_0(2)$};
\node (5)[emptybox] at (1,5) {$\Gamma(2)$};
\node (6)[emptybox] at (2,5) {$\Gamma_0(4)$};
\node (7)[emptybox] at (3,5) {$\mathbf{4C^0}$};
\node (8)[emptybox] at (-2,6) {$\mathbf{4D^0}$};
\node (9)[emptybox] at (0,7) {$\mathbf{4F^0}$};
\node (10)[emptybox] at (2,7) {$\Gamma_0(4) \cap \Gamma(2)$};
\node (11)[emptybox] at (0,9) {$\Gamma(4)$};
\draw (1) -- node[anchor=east] {4}(2);
\draw (1) -- node[anchor=east] {2}(3);
\draw (1) -- node[anchor=east] {3}(4);
\draw (2) -- node[anchor=east] {2}(8);
\draw (2) -- node[anchor=east] {3}(9);
\draw (3) -- node[anchor=east] {4}(8);
\draw (3) -- node[anchor=east] {3}(5);
\draw (4) -- node[anchor=east] {2}(5);
\draw (4) -- node[anchor=east] {2}(6);
\draw (4) -- node[anchor=east] {2}(7);
\draw (5) -- node[anchor=east] {2}(9);
\draw (5) -- node[anchor=east] {2}(10);
\draw (6) -- node[anchor=east] {2}(10);
\draw (7) -- node[anchor=east] {2}(10);
\draw (8) -- node[anchor=east] {3}(11);
\draw (9) -- node[anchor=east] {2}(11);
\draw (10) -- node[anchor=east] {2}(11);
\end{tikzpicture}
    \begin{tabular}{|c|c|c|}
    \hline
         c.c. of $\Gamma$ & $\#$ c.c. & c.c. of Group  \\ \hline
         $\mathrm{PSL}(2,\mathbb{Z})$ & 1 & $\mathfrak{S}_4 = \langle (abcd) , (ab)\rangle$  \\ \hline
         $\mathbf{4A^0}$ & 4 & $\mathfrak{S}_3 =\langle (abc) , (ab)\rangle$    \\ \hline
         $\Gamma^2$ & 1 & $\mathfrak{A}_4=\langle (abc) , (ab)(cd) \rangle$    \\ \hline
         $\Gamma_0(2)$ & 3 & $D_8 =\langle (abcd) , (ac) \rangle$   \\ \hline
         $\Gamma(2)$  & 1 &  $\mathbb{Z}_2^2 =\{1, (ab) (cd) , (ac) (bd) , (ad) (bc) \}$   \\ \hline
         $\Gamma_0(4)$ & 3 &  $\mathbb{Z}_2^2=\langle (ab) (cd) \rangle$  \\ \hline
         $\mathbf{4C^0}$ & 3 &  $\mathbb{Z}_4 =\langle (abcd) \rangle$   \\ \hline
         $\mathbf{4D^0}$ & 4 &  $\mathbb{Z}_3 =\langle (abc) \rangle$  \\ \hline
         $\mathbf{4F^0}$ & 6 &  $\mathbb{Z}_2=\langle (ab) \rangle$   \\ \hline
         $\Gamma_0(4) \cap \Gamma(2)$ & 3 &  $\mathbb{Z}_2=\langle (ab)(cd) \rangle$   \\ \hline
         $\Gamma(4)$ & 1 &  $1$   \\\hline
    \end{tabular}
\caption{Diagram of subgroups of $\mathfrak{S}_4$, and equivalently of congruence subgroups that contain $\Gamma(4)$. The symbols are defined in the table below, where in each class, we also indicate the number of conjugates. We use the nomenclature of \cite{cummins} for congruence subgroups. }
\label{fullDiagsu4}
\end{center}
\end{figure}
For the singlet, the duality group is trivial. 
The characteristic polynomial $P^{\mathfrak{su}(4)}(W)$ defined in terms of the seven 
extremal values of the superpotential is equal to
\begin{equation}
P^{\mathfrak{su}(4)}(W) =\left( W^6 -90 E_4 W^4 -540 E_6
   W^3 -1215 E_4^2 W^2-972 E_4 E_6 W +
108 \left(E_4^3-E_6^2\right)  \right) W  \, . 
\end{equation}
The sextuplet corresponds to the irreducible factor $P^{\mathfrak{su}(4)}(W)/W$.

Our method generalizes to higher rank gauge theories, although the computation complexity grows rapidly. As an illustration, for the smallest $N$ such that $\Gamma_{\textrm{Galois}}(N) \neq \Gamma(N)$, namely $N=8$, we obtain one phase characterized by the order $96$ Galois group
\begin{equation}
G_{\mathfrak{su}(8)} \cong \left( \mathbb{Z}_4 \times \mathbb{Z}_4 \right) \rtimes \mathfrak{S}_3 \, . 
\end{equation}

\section{\texorpdfstring{The $\mathfrak{so}(8)$ Theory}{}}
\label{SectionSo8}

In the $\mathfrak{so}(8)$ ${\cal N}=1^\ast$ theory, more of the power of our framework will come to bare. We will see that in this case, the square root $\sqrt{D}$ of the discriminant of the characteristic polynomial has branch cuts.
Relatedly, the  field extension lies beyond the theory of modular forms for congruent subgroups. The characteristic polynomial will provide us with a full analytic understanding of the vector valued modular objects described numerically in \cite{Bourget:2015cza}.

In \cite{Bourget:2015cza}, we determined the extrema of the $\mathfrak{so}(8)$ exact effective superpotential. We found twenty different extremal values that split into multiplets of one, three, four and twelve extrema -- this last multiplet is called the duodecuplet. The smaller multiplets can be discussed in analogy with previous examples -- the monodromy group is generated by the duality group. The duodecuplet however exhibits new features. The extrema in the duodecuplet were determined numerically in \cite{Bourget:2015cza}, and we computed a Fourier expansion for the value of the superpotential at small absolute value of $q$ (amongst various other properties). We use this Fourier expansion to determine the characteristic polynomial with coefficients in the space of modular forms of $\mathrm{PSL}(2,\mathbb{Z})$.

The characteristic polynomial $P^{\textrm{duod}}(W)$ has degree twelve, and the coefficient of $W^k$ therefore has weight $24-2k$. The dimension of the relevant spaces of modular forms, obtained from (\ref{gradeddimm}), are
\begin{equation}
    \begin{array}{|c|ccccccccccccc|}
    \hline 
        \textrm{Weight} & 0 & 2 & 4 & 6 & 8 & 10 & 12 & 14 & 16 & 18 & 20 & 22 & 24  \\
        \hline
        \textrm{Dimension } & 1 & 0 & 1 & 1 & 1 & 1 & 2 & 1 & 2 & 2 & 2 & 2 & 3 \\
        \hline
    \end{array}
\end{equation}
In \cite{Bourget:2015cza}, we determined more than enough\footnote{Each additional coefficient can be considered as a check of the validity of this computation. We have several dozens such checks.} terms in the Fourier $q$-expansion to identify all these coefficients. The resulting polynomial is 
\begin{eqnarray}
\label{CharacPolyDuodecuplet}
 P^{\textrm{duod}}(W) &=& W^{12} -378 E_4 W^{10}  - 4632 E_6 W^9 - 12951 E_4^2 W^8  \\
    & &  +153144
   E_4 E_6 W^7  +\left(1172340 E_4^3+555708 E_6^2\right) W^6  + 8444160 E_4^2 E_6 W^5  \nonumber \\ 
  & & + \left(5978880 E_4^4+18358272 E_4
   E_6^2\right) W^4  +
   \left(26818560 E_4^3 E_6+16777216 E_6^3\right) W^3   \nonumber \\
  & &   \left(9621504 E_4^5+37748736 E_4^2 E_6^2\right)  W^2 +28311552 E_4^4 E_6 W  + 7077888 E_4^6  \nonumber \, . 
\end{eqnarray}
For this high degree polynomial, we will perform a monodromy analysis only, without attempting to characterize the field extensions or Galois group entirely. We compute the polynomial $\tilde{P}^{\textrm{duod}}$ defined in (\ref{characteristicPolyWeight0}): 
\begin{eqnarray}
\label{CharacPolyDuodecupletZ}
 \tilde{P}^{\textrm{duod}}(Z) &= & \frac{7077888}{k^6}+\frac{28311552 Z}{k^5}+\frac{36864 (1024 k+261) Z^2}{k^5}+\frac{2048 (8192 k+13095) Z^3}{k^4} \nonumber \\
  & & +\frac{6912 (2656 k+865) Z^4}{k^4}+\frac{8444160 Z^5}{k^3}+\frac{12 (46309   k+97695) Z^6}{k^3}\nonumber \\ 
  & & +\frac{153144 Z^7}{k^2}-\frac{12951 Z^8}{k^2}-\frac{4632 Z^9}{k}-\frac{378 Z^{10}}{k}+Z^{12} \, ,  
\end{eqnarray}
where $k(\tau) = \frac{j(\tau)-1728}{j(\tau)}$. Then the discriminant of (\ref{CharacPolyDuodecupletZ}) is 
\begin{eqnarray}
\label{so8discriminant}
  \tilde{D}^{\textrm{duod}} &=& - 2^{166}3^{99} j^{50} (125 j-488095744)^3  \left(14235529   j^3-253052536992000 j^2 \right. \nonumber \\ 
 & & \left. -4425564178752000000 j-24019000623104000000000\right)^2 (j-1728)^{-66} \, . 
\end{eqnarray}
Crucially, we note the presence of the term $(125 j-488095744)^3$ which induces a branch cut in $\sqrt{\tilde{D}^{\textrm{duod}}}$, thus spoiling the modular properties of the roots of the characteristic polynomials by introducing a point of monodromy inside the fundamental domain. Indeed, the equation 
\begin{equation}
   j(\tau) = j_0 := \frac{488095744}{125}. 
\end{equation}
has exactly one solution in the fundamental domain, namely 
\begin{equation}
   \tau_M = i \frac{ _2F_1\left(\frac{1}{6},\frac{5}{6};1;\frac{1}{2}+\frac{2761}{992 \sqrt{31}}\right)}{\, _2F_1\left(\frac{1}{6},\frac{5}{6};1;\frac{1}{2}-\frac{2761}{992
   \sqrt{31}}\right)} \approx 2.4155769875... i \, . 
\end{equation}
This analytic result matches the  prediction made in \cite{Bourget:2015cza}. Let us consider more closely the pair ${W}_i (\tau) $, $ {W}_j (\tau)$ of roots that coincide at $j(\tau) = j_0$. Around the value $j_0$ we have the behavior $({W}_i (q) -  {W}_j (q))^2 \sim (j-j_0)^3$, and then ${W}_i (q) -  {W}_j (q) \sim (j-j_0)^{3/2}$. Thus when $j$ turns  once around $j_0$, the superpotential extrema ${W}_i$ and ${W}_j$ are exchanged, realizing a monodromy of order two in the space of couplings. 

The $\mathfrak{so}(8)$ discriminant (\ref{so8discriminant}) also has a degree three polynomial factor that can vanish. It has one real root and two complex conjugate roots. More precisely, its roots are 
\begin{equation}
\label{solutiondegree3}
   j(\tau) = j_k := \mu^{k}a + \mu^{-k} b + c
\end{equation}
where $\mu = \exp (2 \pi i /3)$ and $k = 1,2,3$ and $a$, $b$ and $c$ are the real numbers 
\begin{eqnarray*}
   a &=& \frac{24000 \sqrt[3]{43606861932191905536849904418-3614096611213342472100883 \sqrt{229}}}{14235529} \\ 
   b &=& \frac{24000 \sqrt[3]{43606861932191905536849904418+3614096611213342472100883 \sqrt{229}}}{14235529} \\
   c &=& \frac{84350845664000}{14235529} \, . 
\end{eqnarray*}
These are order two roots of (\ref{so8discriminant}), meaning that exactly two superpotentials become equal when $ j(\tau) = j_k$. We can write explicit formulas for $\tau$ in the fundamental domain that solve (\ref{solutiondegree3}). Numerical estimations of these exact values are: 
\begin{equation}
    \tau_3 \approx 2.65698 i \qquad \tau_{1,2} \approx \pm 0.43207 + 1.47209 i \, . 
\end{equation}
Around these singular values of $\tau$, the pair of superpotentials that coincide satisfies ${W}_i (q) -  {W}_j (q) \sim (j-j_k)$, and there is no monodromy around these points. This shows that the numerical analysis performed in \cite{Bourget:2015cza} was indeed exhaustive in identifying points of monodromy.

\begin{figure}
\begin{center}
\includegraphics[width=7cm]{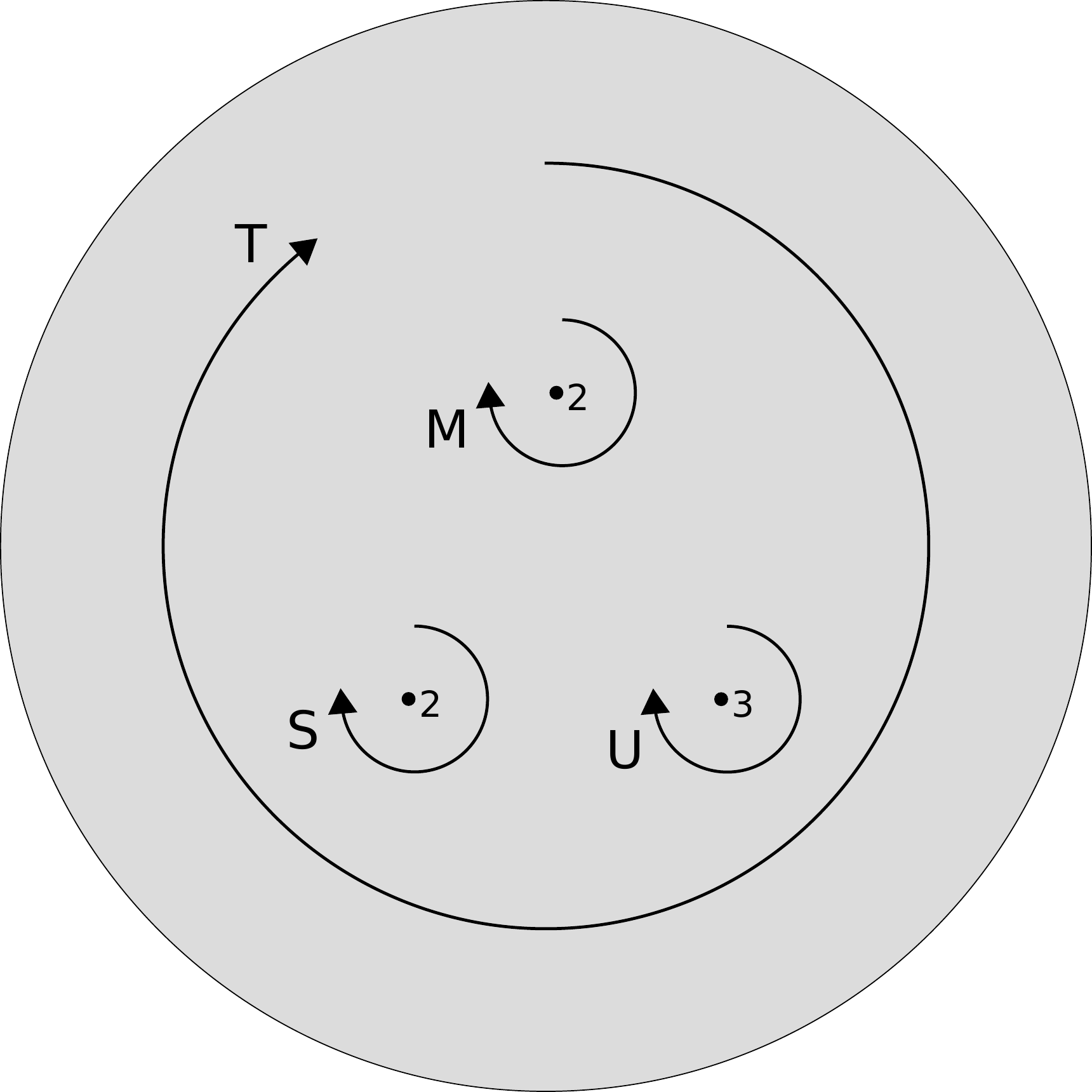}
\end{center}
\caption{The orbifold $\bar{C}$ for the $\mathcal{N}=1^{\ast}$ theory with gauge algebra $\mathfrak{so}(8)$. }
\label{figureCbarso8}
\end{figure}

We can then describe the space $\bar{C}$ in which the complexified gauge coupling lives as an orbifold with the topology of a disk and three elliptic points of orders two, two and three, as pictured in figure \ref{figureCbarso8}. Accordingly the fundamental group is the free product $\pi_1 (\bar{C}) = \mathbb{Z}_2 * \mathbb{Z}_2 * \mathbb{Z}_3$ with generators $M$, $S$ and $U$, which contains a $\mathrm{PSL}(2,\mathbb{Z}) = \mathbb{Z}_2 * \mathbb{Z}_3$ group  as a subgroup. 

In \cite{Bourget:2015cza}, we determined how each generator of $\pi_1 (\bar{C})$ is associated to a permutation of the massive vacua, thus giving the explicit realization of the map $\bar{\sigma}$ (\ref{sigmabar}). Using those results, we can compute e.g. the order of the Galois group. In the duodecuplet, the basic duality transformations were determined to be the following permutations of the twelve vacua:\footnote{Here we shift for convenience the labeling numbers of the twelve vacua of the duodecuplet in \cite{Bourget:2015cza} by minus eight so that they run from 1 to 12. }
\begin{eqnarray}
 \bar{\sigma} (M) &=& (5,8)  \nonumber \\
 \bar{\sigma} (S) &=& (1,2)(3,7)(4,12)(5,8)(6,11)(9,10)  \\ 
 \bar{\sigma} (U) &=& (1,2,7)(3,12,6)(4,5,11)(8,10,9) \, . \nonumber 
\end{eqnarray}
The cardinality of the Galois group $G = \langle \bar{\sigma} (M),\bar{\sigma} (S),\bar{\sigma} (U) \rangle $ generated by these permutations is $|G| = 82944$, which uniquely determines the group to be 
\begin{equation}
\label{cardinalsGroups}
G = \mathfrak{S}_4 \wr   \mathfrak{S}_3
\end{equation}
where $\wr$ denotes the regular wreath product. 

We can also introduce the transformation $T$ which refers to the shift of the $\theta$-angle at weak coupling, or equivalently the monodromy around the point $\tau = i \infty$. This transformation should be identified with the product $ MSU$, as represented in figure \ref{figureCbarso8}, so that $\bar{\sigma} (T) = \bar{\sigma} (MSU) = (2,3,4,5,6,7)(8,9)(11,12)$. It is clear from the figure, and it can be explicitly checked, that those $S$ and $T$ dualities do not satisfy the defining relation of the modular group, $(ST)^3 \neq 1$, as discussed in \cite{Bourget:2015cza}. This explains why the appearance of the additional monodromy $M$ spoils irremediably all modular properties in a given vacuum of the duodecuplet with respect to these $S$ and $T$ actions, in contrast with the analysis of section \ref{SectionSun} where there was a residual modular covariance. As we mentioned previously, the modular group does appear as a subgroup of $\pi_1 (\bar{C})$, although with a different interpretation regarding the complexified coupling constant. For instance, defining $T'=MT=SU$ we have $(ST')^3=U^3=1$. This opens the possibility that the actions of the permutations $S$ and $T'$ be trivial for some congruence subgroup. If this is the case, there exists an integer $N$ such that all the generators of $\Gamma(N)$ are trivial in this representation. Looking at small values of $N$, we observe that this situation indeed occurs for $N=8$ (and not below), as can be checked by expressing the 33 generators of $\Gamma(8)$ using the Farey algorithm. It should be kept in mind though that it is the $T$-transformation that determines the properties of the $q$-expansion at $\tau = i \infty$.

\subsubsection*{Conclusion}

We understand the modular covariant vectors as roots of a characteristic polynomial with coefficients that are modular forms. This provides an analytic understanding of all the integers appearing in the Fourier decomposition of the extrema generated numerically in \cite{Bourget:2015cza}. The superpotential values are modular algebraic, i.e. the superpotential values live in non-analytical extensions of the field of weight two modular forms. More generally,  observables in duality covariant quantum field theory invoke non-analytic modular vectors that live in extensions of the graded ring of modular forms.

In summary, we have realized the two modular group breaking patterns previewed in our  introduction.  The next section will serve to illustrate that it is the full breaking pattern that is generic.

\section{\texorpdfstring{Mass Deformed  ${\cal N}=2$ Superconformal QCD}{}}
\label{SectionQCD}

In this section, we focus on the $\mathfrak{su}(N_c)$, ${\cal N}=2$ theory with $N_f=2 N_c$ massive quarks. Again, we analyze the characteristic polynomial and the permutation group of its roots and its relation to the duality group. We concentrate on the  case of two colors. The Seiberg-Witten analysis for this theory was performed in \cite{Seiberg:1994aj} for $N_c=2$, and in \cite{Argyres:1996eh} for $N_c>2$. The integrable system associated to these theories is the $XXX$ spin chain \cite{Gorsky:1997jq}. Moreover, this integrable system is equivalent to the Gaudin integrable system \cite{Mironov:2012ba}.

We break ${\cal N}=2$ supersymmetry to ${\cal N}=1$ by adding a mass term for the chiral multiplets to the Lagrangian. In other words, we add a quadratic superpotential ${W}(\Phi) = \frac{\mu}{2} \Phi^2$ for the adjoint chiral field $\Phi$. A semi-classical analysis gives the number of massive vacua 
\begin{equation}
    \sum\limits_{r=0}^{N_c -1} (N_c-r) \binom{2N_c}{r} \, . 
\end{equation}
In the literature, there are few results on matching the semi-classical predictions with the results of analyzing the integrable system description of the quantum effective action. In \cite{Hollowood:2004dc}, the theory with $N_c=2$ was studied using the $XXX$ spin chain prescription and the number of vacua and their semi-classical behavior was reproduced from the integrable system.\footnote{We have studied the $XXX$ spin chain prescription of the massive vacua for $N_c=3,4$. The expectation from semi-classics is that the number of vacua is respectively $30$ and $140$. We found the same number of extremal values of the superpotential through a random numerical search algorithm, run for a number of hours on a desktop computer. The technique follows the analysis in \cite{Hollowood:2004dc}, with more variables, and reverting to numerics for determining the zeroes of the resulting high degree polynomials.}

Our main interest is in the monodromies of the massive vacua that arise after breaking ${\cal N}=2$ to ${\cal N}=1$ by mass deformation, in the case of the gauge algebra $\mathfrak{su}(2)$. We denote the four ${\cal N}=2$ masses of the fundamental chiral multiplets by $m_{1,2,3,4}$. In this theory, there are generically six massive vacua. Two vacua arise from the confined dynamics in the $\mathfrak{su}(2)$ low-energy theory when the gauge group is unbroken and $\langle \Phi \rangle = 0$, and the four other vacua are Higgs vacua, associated to the four expectation values, $\langle \Phi \rangle  \propto m_{1,2,3,4} \mathrm{diag}(1,-1)$. We wish to describe the quantum effective superpotential that captures these six vacua, and the permutation of massive vacua under monodromy.

\subsection{The Gaudin Integrable System}

In order to more easily capture the duality action on the ${\cal N}=1$ massive vacua, it is natural to start from the description of the ${\cal N}=2$ curves adapted to these dualities \cite{Gaiotto:2009we}. These curves in turn are more easily interpreted in terms of the Gaudin integrable system. (See also e.g. \cite{Bonelli:2013pva}.) The Gaudin integrable system at hand is a four-punctured two-sphere $S$ with an $\mathfrak{sl}(2)$ valued Hitchin Higgs field $\Phi$. Up to a conformal transformation, the punctures can be located at $z_0=0$, $z_1=1$, $z_\lambda=\lambda$ and $z_\infty=\infty$. The Seiberg-Witten curve $\Sigma$ is a double cover of the sphere $S$ with four branch points at the four punctures. The curve $\Sigma$ is a torus with modular parameter $\tau$ which is related to the cross ratio, or modular $\lambda$-function by 
 \begin{equation}
     \lambda(\tau) = 16 e^{\pi i \tau} \prod\limits_{n=1}^{\infty} \left( \frac{1+e^{2 \pi i n \tau}}{1+e^{2 \pi i \tau ({n- 1/2})}}\right)^8 = \frac{\theta_2^4(\tau)}{\theta_3^4(\tau)} \, . 
 \end{equation}
This relation implies that the (infrared) duality transformation $\tau \rightarrow -1/ \tau$ corresponds to the transformation of the (UV) parameter $\lambda \rightarrow 1-\lambda$, and that the $T$-duality operation $\tau \rightarrow  \tau +1$ corresponds to $\lambda \rightarrow \frac{\lambda}{\lambda-1}$. The cross ratio $\lambda$ is invariant under $T^2 : \tau \rightarrow \tau +2$. In fact, it is invariant under the subgroup $\Gamma(2) \subset \mathrm{SL}(2,\mathbb{Z})$.
 
As before, we are interested in the map from the fundamental group of the space of couplings to the permutation group of the six massive vacua. Let us first describe the space of couplings. It can be parametrized by 
\begin{equation}
   (\lambda , m_i) \in C = \left( \mathbb{C}- \{0,1\} \right) \times \left( \bar{\mathbb{C}} \right)^4 \, , 
\end{equation}
where we have excluded the cusp values $0,1,\infty$ for the cross ratio $\lambda$. See table \ref{tableCuspsElliptic} for a reminder of the cusp and elliptic points and the associated values of the cross ratio. We still have to quotient by the dualities, which provide an equivalence relation: 
\begin{equation}
 (\lambda , m_{0,1,\lambda,\infty} ) \sim  (1-\lambda , m_{1,0,\lambda,\infty}  ) \sim  (\frac{1}{\lambda} , m_{0,\lambda,1,\infty} )  \, . 
\end{equation}
The group generated by those dualities is isomorphic to the permutation group $\mathfrak{S}_3$, and it acts on $C$ as indicated above. The parameter space is then the orbifold quotient 
\begin{equation}
\bar{C} = C  / \mathfrak{S}_3 \, . 
\end{equation}
The fundamental group includes generators of infinite order corresponding to looping the coupling around two of the special points $\lambda \in \{ 0,1, \infty \}$, as well as generators of order two around the points fixed by a non-trivial element of $\mathfrak{S}_3$. 

\begin{table}[t]
    \centering
\begin{tabular}{|c|c|c|c|}
\hline 
   Elliptic point of order 2  & $\tau \in \mathrm{SL}(2, \mathbb{Z}) \cdot i$ & $j = 1728$ & $\lambda = -1,\frac{1}{2} , 2$ \\ \hline
   Elliptic point of order 3  & $\tau \in \mathrm{SL}(2, \mathbb{Z}) \cdot e^{2 \pi i /3}$ & $j = 0$ & $\lambda = e^{2 \pi i /3} , e^{-2 \pi i /3}$ \\ \hline
   Cusps  & $\tau \in \mathrm{SL}(2, \mathbb{Z}) \cdot i \infty = \{i \infty\} \cup \mathbb{Q}$ & $j = \infty$ & $\lambda = 0,1,\infty$ \\ \hline
\end{tabular}
    \caption{Cusps and elliptic points}
    \label{tableCuspsElliptic}
\end{table}

\subsection{The Analysis of the Integrable System Extrema}

\label{polynomialsSQCDstar}

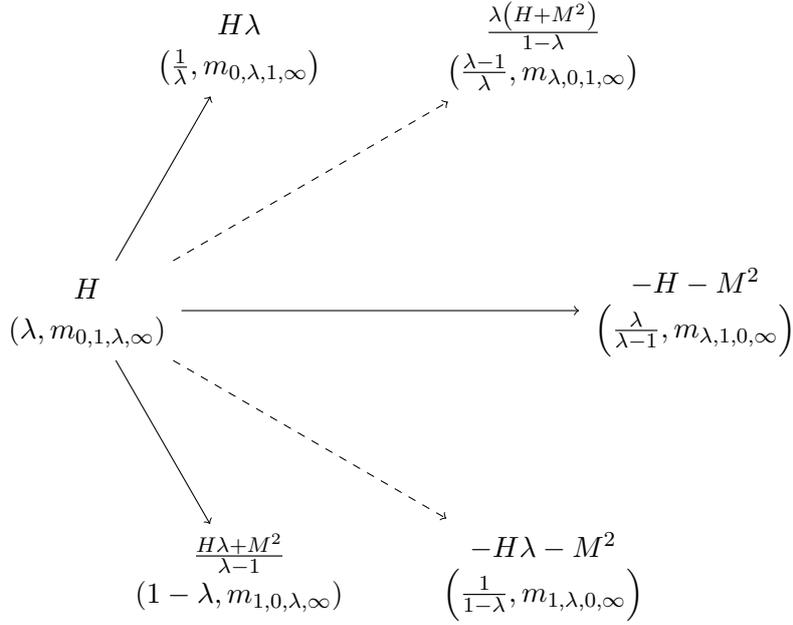
\begin{figure}
\begin{center}
\begin{tikzpicture}[node distance=2cm]
 \node (1) at  (-4,0)  {$\begin{array}{c}
   H   \\ \left( \lambda  , m_{0,1,\lambda,\infty} \right)
 \end{array}$};
 \node (2) at (-2,3.5)  {$\begin{array}{c}
 H \lambda  \\ \left( \frac{1}{\lambda } , m_{0,\lambda,1,\infty} \right)
 \end{array}$};
 \node (3) at (2,3.5) {$\begin{array}{c}
   \frac{\lambda  \left(H+M^2\right)}{1-\lambda }  \\ \left( \frac{\lambda -1}{\lambda } , m_{\lambda,0,1,\infty} \right)
 \end{array}$};
 \node (4) at  (4,0) {$\begin{array}{c}
  -H-M^2  \\ \left( \frac{\lambda }{\lambda -1} , m_{\lambda,1,0,\infty} \right)
 \end{array}$};
 \node (5) at  (2,-3.5){$\begin{array}{c}
   -H \lambda -M^2  \\ \left( \frac{1}{1-\lambda }  , m_{1,\lambda,0,\infty} \right)
 \end{array}$};
 \node (6) at   (-2,-3.5) {$\begin{array}{c}
  \frac{H \lambda +M^2}{\lambda -1}   \\ \left( 1- \lambda  , m_{1,0,\lambda,\infty} \right)
 \end{array}$};
\draw [->] (1) -- (2);
\draw [dashed,->] (1) -- (3);
\draw [->] (1) -- (4); 
\draw [dashed,->] (1) -- (5);
\draw [->] (1) -- (6); 
\end{tikzpicture}
\caption{ Action of the orbifold group $\mathfrak{S}_3$ on the $\lambda$, the masses and the Hamiltonian $H = H_{2,0}$. The plain arrows correspond to the transpositions, also presented in table \ref{tableFixedPoints}, while the dashed arrows represent the three-cycles. To each fixed point of the $\mathfrak{S}_3$ action on the parameter space is associated a monodromy. }
\label{permutationsHamil}
\end{center}
 \end{figure}

The Gaudin integrable model has rank one. We can expect the superpotential
to be determined in terms of the quadratic Hamiltonians of the Gaudin integrable system. 
Let's describe the Hamiltonians in some detail.
We introduce the Gaudin integrable system  through the Higgs field $\Phi$ of the Hitchin system
\begin{eqnarray}
\Phi(z) &=& \sum_{c=0,1,\lambda,\infty} \frac{A^c}{z-z_c} 
\, ,
\end{eqnarray}
where $A^c$ are $\mathfrak{sl}(2)$ matrices, and the adjoint $\mathfrak{sl}(2)$ action
is gauged. 
We moreover impose the 
constraint:
\begin{eqnarray}
\sum_c A^c &=& 0 \, .
\label{sumconstraint}
\end{eqnarray}
We further demand that these matrices lie on a particular orbit, 
\begin{eqnarray}
\mathrm{tr} \, A_i^2 &=& m_i^2 \nonumber \\
\mathrm{tr} \, (A_0+A_1+A_\lambda)^2 &=& m_\infty^2  \, .
\label{Constraints}
\end{eqnarray}
Combining these constraints yields  
\begin{equation}
\label{definitionM}
 \mathrm{tr}(A_0 A_1 + A_1 A_\lambda+A_\lambda A_0) = \frac{1}{2} (m_\infty^2-m_0^2-m_1^2-m_\lambda^2) :=  M^2 \, . 
\end{equation}
The model has five parameters, namely one that defines the geometry of the punctured curve 
(i.e the gauge coupling) and four (masses) that specify the double pole residues of the quadratic differential $\phi_2$ \cite{Gaiotto:2009we}
\begin{equation}
    \phi_2 = \mathrm{tr} \, \left[ \Phi(z)^2 \right] \mathrm{d}z^2 \, . 
\end{equation}
The quadratic Hamiltonians (which are residues of $\mathrm{tr} \, \Phi^2$ at the punctures) are, for $c=0,1,\lambda$, 
\begin{eqnarray}
H_{2,c} &=& \sum_{d \neq c} \frac{\mathrm{tr}(A^c A^d)}{z_c-z_d} \, .
\label{3hamiltonians}
\end{eqnarray}
We will place the fourth singularity  at $z = \infty$. A good coordinates at infinity is $\tilde{z}=\frac{1}{z}$, so that $\mathrm{d}z^2 = \tilde{z}^{-4} \mathrm{d}\tilde{z}^2$, and we have the singularity structure
\begin{equation}
    \phi_2 = \frac{m_{\infty}^2}{\tilde{z}^2} \mathrm{d}\tilde{z}^2 +  \frac{H_{2,\infty}}{\tilde{z}}  \mathrm{d}\tilde{z}^2 + \textrm{finite terms}  \, , 
\end{equation}
with 
\begin{equation}
\label{4thhamiltonian}
    H_{2,\infty} =  \sum\limits_{c , d = 0,1,\lambda} (z_c+z_d) \mathrm{tr} \, ( A^c A^d) \, . 
\end{equation}
While the second order residues at the punctures are specified by our boundary conditions, the first order residues give us four quadratic Hamiltonians (\ref{3hamiltonians}) and (\ref{4thhamiltonian}). A crucial point is that extremizing one of them is equivalent to extremizing any other, up to an affine transformation. This leads to the duality covariance of the ${\cal N}=1$ theory. We will concentrate on the three Hamiltonians (\ref{3hamiltonians}) and the duality transformations implemented by exchanging the points $0,1,\lambda$.
The sum of the three Hamiltonians (\ref{3hamiltonians}) is zero. Upon the exchange of the points $0$ and $1$, we have the interchange of Hamiltonians $H_{2,0} \leftrightarrow H_{2,1}$, and under $1 \leftrightarrow \lambda$ we have $H_{2,\lambda} \leftrightarrow \lambda H_{2,1}$. Under the constraints (\ref{Constraints}), we note the affine relations: 
\begin{equation}
\label{shiftAndScale}
H \equiv H_{2,0} = (\lambda-1) H_{2,\lambda} - M^2 
    = \frac{1-\lambda}{\lambda} H_{2,1} -\frac{M^2}{\lambda} \, .
\end{equation}
In appendix \ref{explicitparametrization}, we introduce an explicit parametrization of the Gaudin integrable system, and compute the equations governing the six massive vacua and their superpotential values.
\label{bulkGaudinexplicit}
We introduce a Lagrange parameter $\alpha$ for the constraint (\ref{sumconstraint}), and after some calculation (see appendix \ref{explicitparametrization}) the extremization procedure reduces to solving a sixth degree polynomial in this variable, namely 
\begin{eqnarray}
\label{polynomial6thorder}
\alpha^2 \left((\alpha-1)^2 \lambda \left(m_\infty^2 (\alpha \lambda-1)^2+m_1^2 (\lambda-1)\right) -m_\lambda^2
   (\lambda-1) (\alpha \lambda-1)^2\right) & & \\ 
-(\alpha-1)^2 m_0^2 (\alpha \lambda-1)^2 &=& 0  \, . \nonumber 
\end{eqnarray}
We can write down the Hamiltonians at the extrema as a function of $\alpha$ alone. For instance for $H \equiv  H_{2,0}$ we find
\begin{equation}
\label{HamiltonianH20alpha}
    H = \frac{(\alpha-1)^2 m_0^2 (\alpha \lambda-1)^2 (\alpha \lambda+\alpha-2)-\alpha^3 (\lambda-1) \left(m_\lambda^2 (\alpha
   \lambda-1)^2-(\alpha-1)^2 m_1^2 \lambda^2\right)}{2 (\alpha-1)^2 \alpha \lambda (\alpha \lambda-1)^2} \, . 
\end{equation}
The value of $H$ in each vacuum is a function of $\lambda$ and the masses $m_{0,1,\lambda,\infty}$ that we call 
\begin{equation}
H_i (\lambda , m_{0,1,\lambda,\infty}) = H (\alpha_i ( \lambda , m_{0,1,\lambda,\infty}), \lambda , m_{0,1,\lambda,\infty}) \, , 
\end{equation}
where the $\alpha_i ( \lambda , m_{0,1,\lambda,\infty})$ are the roots of the polynomial (\ref{polynomial6thorder}). 
We can define a characteristic polynomial
\begin{equation}
\label{polyH}
P_H (H ; \lambda , m_{0,1,\lambda,\infty}) = \prod\limits_{i=1}^6 (H - H_i (\lambda , m_{0,1,\lambda,\infty})) \, . 
\end{equation}
The coefficients of this polynomial are invariant under any permutation acting on the index $i$, so they depend only on the coefficients of the polynomial (\ref{polynomial6thorder}) and can be computed without knowing the roots $H_i (\lambda , m_{0,1,\lambda,\infty})$. Details about this computation and the resulting polynomial $P_H$ are presented in appendix \ref{appendixPolynomials}. With this polynomial at hand, we can observe the action of various dualities. Firstly, the fact that $P_H$ depends only on the function $\lambda (\tau)$ and not directly on $\tau$ shows that the polynomial $P_H$ is invariant under $\Gamma (2)$. Secondly, the polynomial $P_H$ is invariant under transformations such as $\tau \rightarrow \tau +1$ or $\tau \rightarrow -1/\tau$ when combined with the appropriate permutation of the masses. More precisely, all  dualities are obtained from the two identities: 
\begin{eqnarray}
  P_H (H ; \lambda , m_{0,1,\lambda,\infty}) &=& \left(\frac{1}{\lambda} \right) ^6 P_H \left( \lambda H ; \frac{1}{\lambda} , m_{0,\lambda,1,\infty} \right) \, ,  \\
  &=& \left(\frac{\lambda -1}{\lambda} \right) ^6 P_H \left( \frac{\lambda H + M^2}{\lambda -1} ;1-\lambda , m_{1,0,\lambda,\infty} \right) \, . \nonumber
\end{eqnarray}
This shows that knowing a root of the polynomial $P(H ; \lambda , m_{0,1,\lambda,\infty})$ is equivalent to knowing the root of the polynomials obtained after duality as illustrated in figure \ref{permutationsHamil}. A vacuum is therefore characterized by the six values represented in  figure \ref{permutationsHamil}, or equivalently by the six symmetric functions $\mathcal{O}_k$ of these values. The sum $\mathcal{O}_1$ evaluates to $-3 M^2$ and is independent of $\lambda$ and $H$. But the other observables $\mathcal{O}_k$ for $k=2,3,4,5,6$ are non-trivial and characterize the vacua entirely. We have for instance
\begin{equation}
    \mathcal{O}_2 = \frac{-H^2 (\lambda^2 - \lambda +1)^2-H \left(\lambda ^3+1\right) M^2+(\lambda  (3 \lambda -7)+3) M^4}{(\lambda -1)^2} \, ,
\end{equation}
and for generic masses, this operator is generic in the sense of section \ref{section1}. By construction, the operators $\mathcal{O}_k$ are defined on the space $\bar{C}$, and the dualities can be reformulated in terms of the Galois theory associated to their characteristic polynomials $P_{\mathcal{O}_k}$, which we can compute explicitly from the polynomial $P_H$ (\ref{polyH}) -- see appendix \ref{appendixPolynomials} for the result. 

 \subsection{Dualities and Monodromies}

Armed with those polynomials, we can read how the six massive vacua transform along loops in $\bar{C}$, and construct the map $\bar{\sigma}$. The fundamental group $\pi_1 (\bar{C})$ contains four types of elements: 
\begin{enumerate}
    \item elements associated to the monodromies around the excised points $\lambda = 0 , 1 , \infty$, 
    \item elements of order two, associated to the three transpositions in $\mathfrak{S}_3$, 
    \item elements of order three associated to the two three-cycles of $\mathfrak{S}_3$,
    \item cycles around additional points of monodromy, if any. 
\end{enumerate}
Since the three-cycles are the products of two transpositions, it is enough to describe the image by $\bar{\sigma}$ of the elements of type 1 and type 2. The relevant data concerning the corresponding fixed points is summarized in table \ref{tableFixedPoints}. Using the global symmetry of the problem, we note that it is sufficient to focus on one transposition, and to study the permutation of the vacua around the two fixed points in the $\lambda$ space -- one is an orbifold point, the other is one of the excised points, as summarized in the table.  

\begin{table}[t]
    \centering
\begin{tabular}{|c|c|c|c|}
\hline 
    Transposition & Equation & Orbifold fixed point & Excised fixed point \\
    \hline
   $0 \leftrightarrow 1$  & $\lambda = 1- \lambda $ & $\lambda = \frac{1}{2}$ & $\lambda = \infty$ \\
   $1 \leftrightarrow \lambda$  & $\lambda = \frac{1}{\lambda } $ & $\lambda = -1$ & $\lambda = 1$ \\
   $0 \leftrightarrow \lambda$  & $\lambda = \frac{\lambda}{\lambda -1} $ & $\lambda = 2$ & $\lambda = 0$ \\ \hline
\end{tabular}
    \caption{For each transposition in $\mathfrak{S}_3$, we give the fixed point equation and its two solutions. }
    \label{tableFixedPoints}
\end{table}

We focus on the third line of the table. Let us consider first the excised fixed point, namely $\lambda = 0$. The fact that this point is not in $C$ is reflected in the fact that the polynomials $P_{\mathcal{O}_2}$ is singular in the limit $\lambda \rightarrow 0$. However, we can define the rescaled polynomial $\tilde{P}_{\mathcal{O}_2} (\mathcal{O}_2 ; \lambda , m_{0,1,\lambda,\infty})= \lambda^{12}P_{\mathcal{O}_2} (\lambda^{-2} \mathcal{O}_2 ; \lambda , m_{0,1,\lambda,\infty}) $ which is regular in this limit. Plugging in $\lambda = 0$, the polynomial $\tilde{P}_{\mathcal{O}_2} (\mathcal{O}_2 ; 0 , m_{0,1,\lambda,\infty})$ 
factorizes and the roots can be found analytically. Perturbing those roots, we  obtain a small $\lambda$ expansion for the values of $\mathcal{O}_2$ in the six vacua after scaling back to the original polynomial: 
\begin{eqnarray}
\label{expansionsO2}
 \left( \mathcal{O}_2 \right)_1 &=&  -\frac{m_0^2 m_{\lambda}^2}{\lambda ^2}+\frac{2 m_0^3 m_{\lambda}^2}{\lambda  (m_0-m_{\lambda})} + \dots  \nonumber \\
 \left( \mathcal{O}_2 \right)_2 &=& -\frac{m_0^2 m_{\lambda}^2}{\lambda ^2}+\frac{2 m_0^3 m_{\lambda}^2}{\lambda  (m_0+m_{\lambda})}+ \dots \nonumber \\
  \left( \mathcal{O}_2 \right)_3 &=&-\frac{\left(m_0^2+m_{\lambda}^2\right)^2}{4 \lambda ^2}-\frac{\sqrt{(m_0-m_{\lambda}) (m_0+m_{\lambda})}\left(m_0^2+m_{\lambda}^2\right)^{3/2}}{\lambda   ^{3/2}}+ \dots \nonumber \\
 \left( \mathcal{O}_2 \right)_4 &=& -\frac{\left(m_0^2+m_{\lambda}^2\right)^2}{4 \lambda ^2}+\frac{\sqrt{(m_0-m_{\lambda}) (m_0+m_{\lambda})}\left(m_0^2+m_{\lambda}^2\right)^{3/2}}{\lambda   ^{3/2}}+ \dots\nonumber  \\
 \left( \mathcal{O}_2 \right)_5 &=& \frac{-2 \sqrt{m_1^6 \left(m_0^2+m_1^2+m_{\lambda}^2\right)}-m_1^2 \left(m_0^2+2 m_1^2+m_{\lambda}^2\right)}{\lambda ^2} \\
  & & +\frac{2   \left(m_0^2 m_1^2+\sqrt{m_1^6 \left(m_0^2+m_1^2+m_{\lambda}^2\right)}+m_1^4\right)}{\lambda }+ \dots \nonumber \\
 \left( \mathcal{O}_2 \right)_6 &=& \frac{2 \sqrt{m_1^6 \left(m_0^2+m_1^2+m_{\lambda}^2\right)}-m_1^2 \left(m_0^2+2 m_1^2+m_{\lambda}^2\right)}{\lambda ^2} \nonumber \\
  & & +\frac{2 \left(m_0^2   m_1^2-\sqrt{m_1^6 \left(m_0^2+m_1^2+m_{\lambda}^2\right)}+m_1^4\right)}{\lambda }+ \dots \nonumber 
\end{eqnarray}
We can deduce from these expressions that the vacua labeled by $3$ and $4$ are exchanged under $\lambda \mapsto \frac{\lambda}{\lambda -1}$, $m_0 \leftrightarrow m_{\lambda}$ while the others are left invariant.

We now consider the order two permutation around the orbifold point $\lambda = 2$. In order to be at a fixed point of the orbifold action, we also need to impose $m_0 = m_{\lambda}$. Plugging in those values in the characteristic polynomials for the invariant operators, we find that they factorize as
\begin{equation}
P_{\mathcal{O}_k} (\mathcal{O}_k) = \left( Q_k (\mathcal{O}_k)\right)^2 \, , 
\end{equation}
where $Q_k$ is a degree three polynomial, whose roots are distinct for generic masses. As a consequence, the image by the map $\bar{\sigma}$ of this orbifold cycle is an element of $\mathfrak{S}_2 \times \mathfrak{S}_2 \times \mathfrak{S}_2$. One can show, using either numerical computations or exact perturbation theory, that the element is a product of three transpositions.

Finally, we turn to possible additional points of monodromy. Let us first consider a simple and interesting situation in which  $M=0$ and $m_0=m_1=m_{\lambda} \equiv m$. We then find the discriminant
\begin{eqnarray}
\mathrm{disc} \, P_{\mathcal{O}_2} &=& -\frac{282429536481}{4 (\lambda -1)^{52} \lambda ^{52}} (\lambda -2)^6 (\lambda +1)^6 (2 \lambda -1)^6 \left(\lambda ^2-\lambda +1\right)^{60} \nonumber  \\
& & \times \left(\lambda ^6-3 \lambda ^5+60 \lambda ^4-115 \lambda ^3+60 \lambda ^2-3 \lambda +1 \right)^3 m^{120} \nonumber \\
&=& - 3^{24}2^{-204} m^{120} j^{20} (j-1728)^3 (j+13824)^3  \, . 
\label{discPO2}
\end{eqnarray}
We  used the relation between the $\lambda$ function and the $j$ invariant, 
\begin{equation}
j(\tau) = \frac{256 \left(\lambda ^2-\lambda +1\right)^3}{(1-\lambda )^2 \lambda ^2} \, . 
\end{equation}
Because we concentrated on the mass configuration $m_0=m_1=m_{\lambda}$, the discriminant (\ref{discPO2}) vanishes at the elliptic points and diverges at the cusps, see table \ref{tableCuspsElliptic}. The discriminant also has additional zeros. They correspond to a complexified coupling constant $\tau$ that satisfies $j(\tau) = -13824 $.\footnote{We note that $1728=12^3$ and $13824=24^3$.}
Let us concentrate on the additional root
\begin{equation} 
\lambda_M = \frac{1}{2} \left(1-i \sqrt{3 \left(25+20 \sqrt[3]{2}+16\ 2^{2/3}\right)}\right) \, . 
\end{equation} 
Plugging this value into the polynomial $P_{\mathcal{O}_2}$, we find a double root $\mathcal{O}_2 = 27 m^4$, and we can then compute the evolution of these two roots for $\lambda \neq \lambda_M$ perturbatively, 
\begin{eqnarray}
  m ^{-4} \mathcal{O}_2 &=& 27 +45 i \sqrt{6 \left(12-7 \sqrt[3]{2}-2\ 2^{2/3}\right)} (\lambda -\lambda_M  ) \\
  & & \pm 12 i \sqrt[4]{-4539+4986 \sqrt[3]{2}-1098\ 2^{2/3}} (\lambda -\lambda_M )^{3/2}  + O((\lambda -\lambda_M )^{2}) \, . 
\end{eqnarray}
The non-zero coefficient in front of $ (\lambda -\lambda_M )^{3/2}$ proves that the two vacua are exchanged when $\lambda$ circles once around the additional point of monodromy $\lambda_M$. In this very symmetric configuration we have thus been able to demonstrate the existence of additional points of monodromy. Because of their topological character, these points must survive when going away from this symmetric configuration. We deduce that the phenomenon of additional monodromies is generic. This conclusion is confirmed by numerical computations. 

As a final remark, let us note that we can also connect  the quantum behavior of massive vacua to our semi-classical intuition. The monodromy around the point $\lambda = 0$ can be interpreted as $T$-duality. The expansions (\ref{expansionsO2}) prove that two vacua admit an expansion in $\lambda^{1/2}$, and can be identified with the two confining vacua, while the four other have an expansion in integer powers of $\lambda$ and are the Higgs vacua. 

\subsubsection*{Conclusions}
We conclude that the $SU(2)$ $N_f=4$ theory with ${\cal N}=1$ supersymmetric mass deformation  has a Galois group that includes monodromies that do not descend from ${\cal N}=2$ dualities, and that the superpotential expectation values fully break the duality group. While the duality covariance of the ${\cal N}=2$ theory governs the coefficients of the characteristic polynomials, the choice of massive ${\cal N}=1$ vacuum breaks duality. After mass deformation, it is the Galois permutation group that classifies quantum vacua connected by interpolation in the coupling space. 

\section{Closing Section}
The vacuum and chiral ring structure of ${\cal N}=1$ theories is a rich subject. The mere enumeration of the massive vacua gives rise to intriguingly rich counting problems.  In this paper, we determined the permutation of massive vacua  under completing a loop in the space of couplings. This quest lead us into the Galois theory of field extensions. In the case of the ${\cal N}=1$ theories that inherit duality covariance properties from their ${\cal N}=2$ parents, we argued that the Galois group takes over the role of the duality group in classifying multiplets of massive vacua.

It is clear that many open problems remain. It would be interesting to promote  our results on the permutations of vacua into a study of the full Berry connection associated to ${\cal N}=1$ vacuum states. Tackling the  problem of determining the full chiral ring vacuum expectation values in some low rank examples with duality covariance, in the spirit of \cite{Ferrari:2008rz,Ferrari:2009zh}, should be worthwhile. An efficient derivation of the characteristic polynomials from scratch would also be most welcome. In short, if we promote our generic grasp of the physics of ${\cal N}=1$ theories  to a much more detailed understanding of an ever larger class of theories, we are bound to unearth more beautiful treasures.

\subsubsection*{Acknowledgments}
We would like to acknowledge support from the grant ANR-13-BS05-0001, the EU CIG grant UE-14-GT5LD2013-618459 as well as Asturias Government grant FC-15-GRUPIN14-108 and Spanish Government grant MINECO-16-FPA2015-63667-P. 
It is our pleasure to thank Gaëtan Chenevier for useful explanations and Costas Bachas, Amihay Hanany, Prem Kumar and Diego Rodriguez-Gomez for discussions.

\appendix
\section{Congruence Subgroups, Automorphic Forms and Galois Theory}
\label{appendixCongSub}

In this appendix, we first collect useful facts about congruence subgroups and the associated fields of functions, which are the basic ingredients of the Galois correspondence. We also explain how these fields are related to the graded ring of modular forms, using the encompassing concept of automorphic form. 

\subsection{Congruence Subgroups}
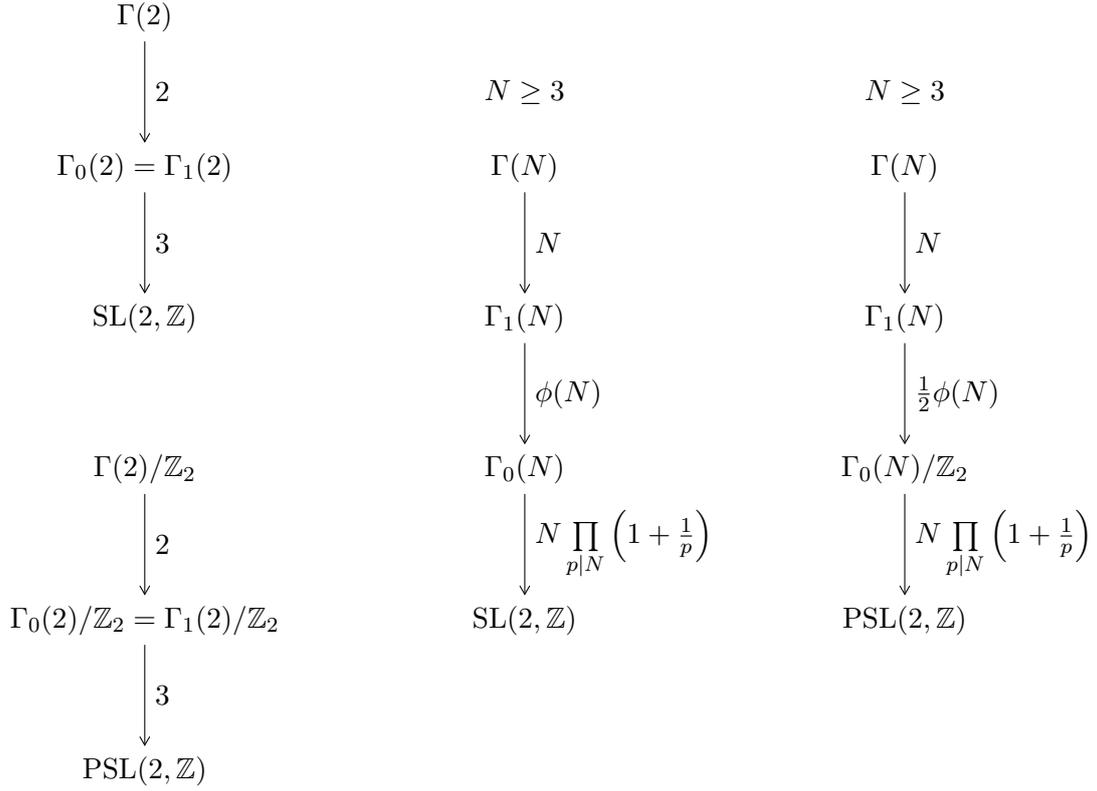
\begin{figure}
\begin{center}
\begin{tikzpicture}[node distance=2cm]
\node (1) [emptybox] {$\mathrm{PSL}(2,\mathbb{Z})$};
\node (2)[emptybox, above of=1] {$\Gamma_0 (2)/\mathbb{Z}_2=\Gamma_1 (2) / \mathbb{Z}_2$};
\node (3)[emptybox, above of=2] {$\Gamma (2) /\mathbb{Z}_2$};
\draw [arrow] (2) -- node[anchor=west] {$3$}(1);
\draw [arrow] (3) -- node[anchor=west] {$2$}(2);
\node (31) [emptybox] at (0,6) {$\mathrm{SL}(2,\mathbb{Z})$};
\node (32)[emptybox, above of=31] {$\Gamma_0 (2)=\Gamma_1 (2)$};
\node (33)[emptybox, above of=32] {$\Gamma (2)$};
\draw [arrow] (32) -- node[anchor=west] {$3$}(31);
\draw [arrow] (33) -- node[anchor=west] {$2$}(32);
\node (11) [emptybox] at (5,2) {$\mathrm{SL}(2,\mathbb{Z})$};
\node (12)[emptybox, above of=11] {$\Gamma_0 (N)$};
\node (13)[emptybox, above of=12] {$\Gamma_1 (N)$};
\node (14)[emptybox, above of=13] {$\Gamma (N)$};
\node (15)[emptybox, above of=14,yshift=-1cm] {$N \geq 3$};
\draw [arrow] (12) -- node[anchor=west] {$N \prod\limits_{p|N} \left( 1+\frac{1}{p}\right)$}(11);
\draw [arrow] (13) -- node[anchor=west] {$\phi (N)$}(12);
\draw [arrow] (14) -- node[anchor=west] {$N$}(13); 
\node (21) [emptybox] at (10,2) {$\mathrm{PSL}(2,\mathbb{Z})$};
\node (22)[emptybox, above of=21] {$\Gamma_0 (N) / \mathbb{Z}_2$};
\node (23)[emptybox, above of=22] {$\Gamma_1 (N)$};
\node (24)[emptybox, above of=23] {$\Gamma (N)$};
\node (25)[emptybox, above of=24,yshift=-1cm] {$N \geq 3$};
\draw [arrow] (22) -- node[anchor=west] {$N \prod\limits_{p|N} \left( 1+\frac{1}{p}\right)$}(21);
\draw [arrow] (23) -- node[anchor=west] {$\frac{1}{2} \phi (N)$}(22);
\draw [arrow] (24) -- node[anchor=west] {$N$}(23); 
\end{tikzpicture} 
\caption{Some congruence subgroups of level $N$  with the indices of the group inclusions. On the right are congruence subgroups of $\mathrm{PSL}(2,\mathbb{Z})$ of level $N$. } 
\label{congSub} 
\end{center} 
\end{figure}

A subgroup of $\mathrm{SL}(2,\mathbb{Z})$ is called a congruence subgroup if there exists an integer $N \geq 1$ such that it contains the group
\begin{equation}
   \Gamma(N) := \left\{  \left(\begin{array}{cc} a & b \\ c & d\end{array}  \right) \in \mathrm{SL}(2,\mathbb{Z}) | b \equiv c \equiv  0  \textrm{ and } a \equiv d \equiv  1 \textrm{ mod } N \right\} \, . 
\end{equation}
Common congruence subgroups are 
\begin{equation}
   \Gamma_0(N) := \left\{  \left(\begin{array}{cc} a & b \\ c & d\end{array}  \right) \in \mathrm{SL}(2,\mathbb{Z}) |  c \equiv  0  \textrm{ mod } N \right\}  \, ,
\end{equation}
and
\begin{equation}
   \Gamma_1(N) := \left\{  \left(\begin{array}{cc} a & b \\ c & d\end{array}  \right) \in \mathrm{SL}(2,\mathbb{Z}) |   c \equiv  0  \textrm{ and } a \equiv d \equiv  1  \textrm{ mod } N \right\}  \, .
\end{equation}
We give the inclusion structure of these groups and the associated indices in figure \ref{congSub}. In particular, the index of $\Gamma(N)$ is
\begin{equation}
   [\mathrm{PSL}(2,\mathbb{Z}) , \Gamma(N)] = \begin{cases} \frac{N^2}{2} \phi (N)  \prod\limits_{p|N} \left( 1+\frac{1}{p}\right) & \textrm{ for } N > 2 \\ 6 & \textrm{ for } N=2 \, , \end{cases}
\end{equation}
where $\phi$ is the Euler totient function. A useful relation is, for all $N \geq 2$, 
\begin{equation}
 [\mathrm{SL}(2,\mathbb{Z}) , \Gamma(N)] = |\mathrm{SL}(2,\mathbb{Z}_N)| = N^3 \prod\limits_{p|N} \left( 1- \frac{1}{p^2}\right) = N^2 \phi (N) \prod\limits_{p|N} \left( 1+ \frac{1}{p}\right) \, . 
\end{equation}

\subsection{Galois Correspondence}

Now that we have described the classical congruence subgroup and their inclusions, we recall how to attach a field to each of them and describe those fields explicitly. 

From any congruence subgroup $\Gamma$ we can construct a compact Riemann surface, called the \emph{modular curve} $\mathbf{X}(\Gamma)$, which is a compactification \cite{shimura,diamond} of $\Gamma \backslash \mathfrak{H}$. We denote by $\mathbb{C}(\mathbf{X}(\Gamma))$ the field of meromorphic functions on this curve. If the genus of $\mathbf{X}(\Gamma)$ is zero, then the field can be generated by a single function (sometimes called the Hauptmodul), and otherwise more generators are needed.\footnote{In fact, two generators always suffice, but the fields can be more easily described using more generators, as we do here for $\mathbb{C}(\mathbf{X}(\Gamma(N)))$. } 

Let us describe these fields for the classical congruence subgroups introduced in the previous subsection. For $(a,b) \in \mathbb{Z}_N^2- \{(0,0)\}$, we define the functions
\begin{equation}
   \mathbf{f}_N^{a,b} (\tau)= \frac{E_4(\tau)}{E_6(\tau)} \wp \left( \frac{a \tau +b}{N} ; \tau \right) \, . 
\end{equation}
These are $\frac{1}{2}(N^2-1)$ (for $N \geq 3$, and three for $N=2$) distinct functions that satisfy $\mathbf{f}_N^{a,b} = \mathbf{f}_N^{-a,-b}$, and the dualities act via $T : \mathbf{f}_N^{a,b} \rightarrow \mathbf{f}_N^{a,a+b}$ and $S : \mathbf{f}_N^{a,b} \rightarrow \mathbf{f}_N^{-b,a}$. 
Define also 
\begin{equation}
   j_N (\tau) = j(N \tau) \, . 
\end{equation}
We then have \cite{diamond}
\begin{eqnarray}
\mathbb{C}(\mathbf{X}(\Gamma_0(N))) &=& \mathbb{C}(j,j_N) \\
\mathbb{C}(\mathbf{X}(\Gamma_1(N))) &=& \mathbb{C}(j,\mathbf{f}_N^{1,0}) \\
\mathbb{C}(\mathbf{X}(\Gamma(N))) \,  &=& \mathbb{C}(j,\mathbf{f}_N^{1,0},\mathbf{f}_N^{0,1}) \, . 
\end{eqnarray}
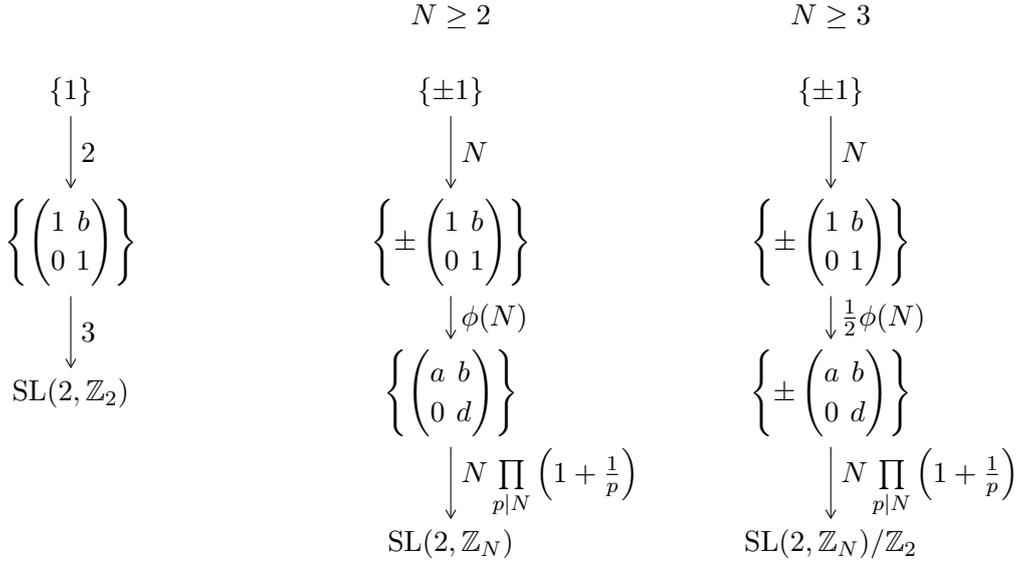
\begin{figure}
\begin{center}
\begin{tikzpicture}[node distance=2cm]
\node (31) [emptybox] at (0,4) {$\mathrm{SL}(2,\mathbb{Z}_2)$};
\node (32)[emptybox, above of=31] {$\left\{  \begin{pmatrix}  1 & b \\ 0 & 1 \end{pmatrix} \right\}$};
\node (33)[emptybox, above of=32] {$\{ 1 \}$};
\draw [arrow] (32) -- node[anchor=west] {$3$}(31);
\draw [arrow] (33) -- node[anchor=west] {$2$}(32);
\node (11) [emptybox] at (5,2) {$ \mathrm{SL}(2,\mathbb{Z}_N)$};
\node (12)[emptybox, above of=11] {$ \left\{  \begin{pmatrix}  a & b \\ 0 & d \end{pmatrix} \right\}$};
\node (13)[emptybox, above of=12] {$\left\{ \pm  \begin{pmatrix}  1 & b \\ 0 & 1 \end{pmatrix} \right\}$};
\node (14)[emptybox, above of=13] {$\{\pm 1\}$};
\node (15)[emptybox, above of=14,yshift=-1cm] {$N \geq 2$};
\draw [arrow] (12) -- node[anchor=west] {$N \prod\limits_{p|N} \left( 1+\frac{1}{p}\right)$}(11);
\draw [arrow] (13) -- node[anchor=west] {$\phi (N)$}(12);
\draw [arrow] (14) -- node[anchor=west] {$N$}(13); 
\node (21) [emptybox] at (10,2) {$ \mathrm{SL}(2,\mathbb{Z}_N) / \mathbb{Z}_2$};
\node (22)[emptybox, above of=21] {$ \left\{ \pm  \begin{pmatrix}  a & b \\ 0 & d \end{pmatrix} \right\}$};
\node (23)[emptybox, above of=22] {$\left\{ \pm  \begin{pmatrix}  1 & b \\ 0 & 1 \end{pmatrix} \right\}$};
\node (24)[emptybox, above of=23] {$\{\pm 1\}$};
\node (25)[emptybox, above of=24,yshift=-1cm] {$N \geq 3$};
\draw [arrow] (22) -- node[anchor=west] {$N \prod\limits_{p|N} \left( 1+\frac{1}{p}\right)$}(21);
\draw [arrow] (23) -- node[anchor=west] {$\frac{1}{2} \phi (N)$}(22);
\draw [arrow] (24) -- node[anchor=west] {$N$}(23); 
\end{tikzpicture} 
\caption{Galois Groups associated to the congruence subgroups of figure \ref{congSub}. The group denoted $\{\pm 1\}$ at the top is the trivial group (of cardinality $1$) inside $ \mathrm{SL}(2,\mathbb{Z}_N) / \mathbb{Z}_2$. The letters $a,b,d$ denote generic elements in $\mathbb{Z}_N$, and all matrices have unit determinant. } 
\label{congSub2} 
\end{center} 
\end{figure} 
The fields of meromorphic functions on the modular curves $\mathbf{X}(\Gamma)$ are fairly well understood. In order to translate this knowledge into information about modular forms, we introduce the unifying concept of automorphic form. A function $f : \mathfrak{H} \rightarrow \hat{\mathbb{C}}$ is an automorphic form of weight $k$ for $\Gamma$, which we denote by $f \in \mathcal{A}_k(\Gamma)$, if 
\begin{enumerate}
    \item $f$ is meromorphic on $\mathfrak{H}$ and at the cusps
    \item $f$ is weight-$k$ invariant under $\Gamma$. 
\end{enumerate}
Note that $f \in \mathcal{A}_k(\Gamma)$ is well defined on $\mathbf{X}(\Gamma)$ when $k=0$, but not when $k \neq 0$ unless $f=0$. The fundamental property is 
\begin{equation}
   \forall f \in \mathcal{A}_k(\Gamma) - \{0\} , \qquad \mathcal{A}_k(\Gamma) =\mathbb{C}(\mathbf{X}(\Gamma)) f \, . 
\end{equation}
Thus knowing $\mathcal{A}_0(\Gamma) =\mathbb{C}(\mathbf{X}(\Gamma)) $ and a non-trivial element of $\mathcal{A}_k(\Gamma)$ gives a full description of $\mathcal{A}_k(\Gamma)$. We recall that if $k\geq 0$ is even, then 
\begin{equation}
(E_6/E_4)^{k/2} \in \mathcal{A}_k(\Gamma) - \{0\}  \, , 
\end{equation}
The modular forms $\mathcal{M}_k(\Gamma)$ are those automorphic forms that are holomorphic on $\mathfrak{H}$ and at the cusps. Applying the Riemann-Roch theorem, one can deduce that this regularity condition restricts $\mathcal{A}_k(\Gamma)$ to a finite-dimensional subspace whose dimension is given by the formula quoted in (\ref{dimringMF}).

\section{\texorpdfstring{Properties of $\mathcal{N}=4$   $\mathfrak{su}(N)$ Theories}{}}
We collect in this appendix two related analyses. The first is an analysis of the space of ${\cal N}=4$
super Yang-Mills theories with $\mathfrak{su}(N)$ gauge algebra. The second part contains proofs of properties
of lattices.
\subsection{\texorpdfstring{The $\mathcal{N}=4$ Theories with $\mathfrak{su}(N)$ Gauge Algebra}{}}
\label{appendixTheoriessuN}

The authors of \cite{Aharony:2013hda} have shown that the number of different $\mathcal{N}=4$ theories with gauge algebra $\mathfrak{su}(N)$ is given by $\sigma_1(N)$, since we have to make a choice of center $\mathbb{Z}_d \in \mathbb{Z}_N$ of the gauge group, and one among $d$ possible tilts in the spectrum of dyonic line operators. The structure of the set of $\mathfrak{su}(N)$ ${\cal N}=4$ (and ${\cal N}=1^\ast$) theories is therefore in bijection with the lattices (\ref{equationLattice}), which are in this context interpreted as the electric and magnetic charges of line operators of the theory.

The space of parameters for a theory of level $d$ is $\Gamma_0(N/d^2) \backslash \mathfrak{H}$. Indeed, a given theory, characterized by a lattice $L_{N,p,k}$, is left invariant under a group of the form $\alpha^{-1} \Gamma_0(N/d^2) \alpha $, which is isomorphic to $\Gamma_0(N/d^2)$. For instance, the $SU(N)$ theory is associated with the lattice $L_{N,1,0}$ which is left invariant by 
$\Gamma_0(N)$.\footnote{The associated lattice $L_{N,1,0}$ contains points of the form $(r,0)$ for $r \in \mathbb{Z}_N$, and therefore acting with $\left(\begin{matrix} a & b \\ c & d \end{matrix} \right) \in \Gamma_0(N)$ gives $(ar,cr) = (r',0)$ with $r' \in \mathbb{Z}_N$ since necessarily $\textrm{gcd}(a,N)=1$. The proof of the statement given in \cite{Aharony:2013hda} is based on the incorrect assumption that the group $\Gamma_0(N)$ is generated by $T$ and $S T^N S$. The statement, however, is correct.} If $N$ is square-free, then all lattices have level one, and we recover the results of \cite{Aharony:2013hda}.

Note however that for the local physics we describe in the rest of the paper, the global aspects we discussed in this appendix will play second fiddle, and we may then simply refer to $\mathrm{PSL}(2,\mathbb{Z})$ as the duality group.

\subsection{Proofs of the Properties of Section \ref{subsecSubgroupsZN2}}
\label{AppendixProofs}

We prove the three properties stated in section \ref{subsecSubgroupsZN2} that are useful in 
appendix \ref{appendixTheoriessuN} as well. We keep the convention that when $p$ is a divisor of $N$, we set $q= \frac{N}{p}$. 
\begin{enumerate}
    \item For a divisor $p$ of $N$, consider the $p$ lattices $L_{N,p,k}$, with $k \in \mathbb{Z}_p$. The action of $T$ on this set of lattices is $k \mapsto k+ q$, so the size of an orbit is the smallest positive solution of $x q \equiv 0$ modulo $p$. This solution satisfies $xq = \mathrm{lcm} \left( p ,q\right)$, so the orbit has length $p / \mathrm{gcd} \left( p , q\right)$. Since all orbits in $\mathbb{Z}_p$ have the same length, we conclude there are $\mathrm{gcd} \left( p , q \right)$ of them. 
    \item We prove the two implications separately. 
    \begin{itemize}
        \item Firstly, let us prove that the level is preserved by any duality operation. We remark that the level of a lattice is equal to the greatest common divisor of the coordinates of all the points in the lattice. Indeed, since $d$ divides $p$, $k$ and $q$ in (\ref{equationLattice}), it divides all the coordinates, and reciprocally a divisor of all the coordinates will divide the coordinates of the points $(p,0)$ and $(k,q)$. Now let $L_{N,p,k}$ be a lattice of level $d$. Left multiplication by a matrix with integer coefficients can not decrease the level. If the matrix is in $\mathrm{SL}(2,\mathbb{Z})$, the operation is invertible, hence the level is preserved. 
    \item Now we prove that conversely, if two lattices have the same level, they are related by a duality. Without loss of generality, we can focus on lattices of level $1$, dividing if necessary by the level of the initial lattice. Let us then consider $L_{N,p,k}$ with $\mathrm{gcd} \left( p ,q ,k \right) = 1$. Under this assumption, there exists an integer\footnote{One such integer $l$ can be constructed explicitly along the lines of the proof of Lemma B.1 in \cite{Bourget:2016yhy}, as the product of all prime numbers that divide $p$ but that don't divide $k$ nor $q$.} $l$ such that $\mathrm{gcd} \left( p , q l +k \right) = 1$. This means that $T^l L_{N,p,k} = L_{N,p,k'}$ with $\mathrm{gcd} (p,k')=1$. Taking the $S$-dual of this lattice, one obtains $S T^l L_{N,p,k} = L_{N,N,k''}  $ for some integer $k''$, and finally $ST^{-k''}S T^l L_{N,p,k} = L_{N,1,0}$. We have proven that any lattice of level $1$ is related by a sequence of dualities to $L_{N,1,0}$, so any two such lattices are related by a duality sequence as well. 
    \end{itemize}
    \item Again we focus on lattices of level $1$. Let $L_{N,p,k}$ be such a lattice. According to the previous paragraph, it can be related by a chain of dualities to $L_{N,1,0}$. Let us denote $\alpha$ the corresponding $\mathrm{SL}(2,\mathbb{Z})$ matrix, so that $L_{N,1,0} = \alpha L_{N,p,k}$. The lattice $L_{N,1,0}$ is left invariant by $\Gamma_0 (N)$, so $L_{N,p,k}$ is left invariant by $\alpha^{-1} \Gamma_0 (N) \alpha$. 
\end{enumerate}

\section{Appendices to Section \ref{SectionQCD}}
We gather a number of technical results relevant to section \ref{SectionQCD}.
\subsection{The Parametrization of the Gaudin Model}
\label{explicitparametrization}
In this subsection, we provide a few technical details of the analysis of the Gaudin integrable system extrema exploited in section \ref{bulkGaudinexplicit}. Firstly, we parameterize the $\mathfrak{sl}(2)$ valued matrices $A_i$ ($i=0,1,\lambda$) using complex numbers $a_i,c_i$ as follows 
\begin{equation}
A_i =   \frac{m_i}{\sqrt{2}}  \left(
\begin{array}{cc}
 2 a_i c_i+1 & 2c_i (a_i c_i+1) \\
 -2 a_i & -2 a_i c_i -1 \\
\end{array}
\right) \, . 
\end{equation}
The conditions $Tr A_i = 0$ and $Tr A_i^2 = m_i^2$ are then automatically enforced. 
The constraint $\sum_c A_c =0$ fixes the  matrix $A_\infty$.
We exploit the gauge freedom of conjugating all three remaining matrices by $\mathrm{SL}(2,C)$ which is of dimension three. Firstly,  we can use conjugation to diagonalize $A_\lambda$, which amounts to setting $c_\lambda=a_\lambda=0$. We can then still conjugate by a diagonal matrix
and put  $A_1$ in the form
\begin{equation}
\left(
\begin{array}{cc}
 2 b+1 & 2 b (b+1) \\
 -2  & -2 b-1 \\
\end{array}
\right) \, ,
\end{equation}
where $a_1=1$ and $c_1=b$. 
To simplify notation further, we also set $a_0=a$, $c_0=c$.  We then define a functional $W_0$ to be extremized, with a Lagrange multiplier $\alpha$ that imposes the last constraint
\begin{eqnarray}
W_0(a,b,c,\alpha) &=& H_{2,0} + \alpha (\sum_{k <j}  Tr (A_k A_j) - M^2) \, .
\label{toextremize}
\end{eqnarray}
The explicit expression reads
\begin{eqnarray}
  \frac{\lambda}{m^2}  W_0 &=& -2 a \left(c (-2 b \alpha \lambda+2 b \lambda-2 \alpha \lambda+\lambda+1)+b (b+1) (\alpha-1) \lambda+c^2 (\alpha-1)
   \lambda\right) \nonumber \\ 
  & & +\lambda (4 b \alpha-2 b-2 c \alpha+2 c+4 \alpha-1)-1 \, . 
\end{eqnarray}
We extremize this system with respect to $a,b,c$ and $\alpha$
and after elimination obtain a sixth order polynomial in
$\alpha$ that governs the positions of the extrema. The extremal values are then given
by evaluating the potential $W_0$ in equation (\ref{toextremize}) at the roots of the sixth
order polynomial. The final equations
are recorded in section \ref{bulkGaudinexplicit}.

\subsection{Explicit Polynomials}
\label{appendixPolynomials}

In section \ref{polynomialsSQCDstar}, in order to construct the polynomials $P_H$ and $P_{\mathcal{O}_k}$ we face the problem of finding a polynomial $Q$ whose roots are the images by some function $f$ of the roots of another polynomial $P = \prod_i (X - \alpha_i)$. If the function $f$ is a ratio of polynomials in $\alpha$, $f(\alpha) = \frac{f_{\textrm{num}} (\alpha)}{f_{\textrm{den}} (\alpha)}$, this problem can be solved without knowing explicitly the roots $\alpha_i$. Indeed, we can consider 
\begin{equation}
Q(X) = \prod\limits_i \left( f_{\textrm{den}} (\alpha) X -  f_{\textrm{num}} (\alpha) \right)
\end{equation}
whose coefficients are symmetric polynomials of the $\alpha_i$. Using a reduction algorithm for symmetric polynomials, these coefficients can be expressed in terms of the elementary symmetric functions in the $\alpha_i$, which in turn can be identified up to a sign to the coefficients of the polynomial $P$. 

We catalogue below explicit expressions for the polynomials that feature in section \ref{polynomialsSQCDstar}, obtained using this algorithm. Although we can compute the polynomials in full generality, for simplicity we provide them here only in the situation where the masses satisfy the constraint $M=0$, or equivalently $m_{\infty}^2 = m_0^2+m_1^2+m_{\lambda}^2$. 

The characteristic polynomial for $H$ is 
\begin{equation}
P_H (H ; \lambda , m_{0,1,\lambda}) = H^6 + \sum\limits_{j=0}^5 r_j (\lambda , m_{0,1,\lambda}) H^j
\end{equation}
where the functions $r_j (\lambda , m_{0,1,\lambda,\infty})$ are 
\begin{eqnarray*}
 r_0 &=& \frac{m_0^2 m_1^2 m_{\lambda}^2}{4 \lambda ^6} \left(3 (\lambda -1)^4 m_1^4 \left(\lambda ^2 m_0^2+m_{\lambda}^2\right)+\left(\lambda ^2 m_0^2+m_{\lambda}^2\right)^3 \right. \\ 
 & & \left. +3 (\lambda   -1)^2 m_1^2 \left(\lambda ^4 m_0^4-7 \lambda ^2 m_0^2 m_{\lambda}^2+m_{\lambda}^4\right)+(\lambda -1)^6 m_1^6\right) \\
 r_1 &=&  -\frac{3 m_0^2 m_1^2 m_{\lambda}^2}{2 \lambda ^5} \left(\lambda ^2 m_0^2+(\lambda -1)^2 m_1^2+m_{\lambda}^2\right) \left((\lambda -2) \lambda ^2 m_0^2+(\lambda -1)^2 (\lambda +1)
   m_1^2+(1-2 \lambda ) m_{\lambda}^2\right) \\
 r_2 &=&  -\frac{1}{4 \lambda ^4} \left(\lambda ^4 m_{0}^6 \left(m_{1}^2+m_{\lambda}^2\right)-\lambda ^2 m_{0}^4 \left(-2 (\lambda -1)^2 m_{1}^4+\left(9 \lambda ^2-26 \lambda +26\right) m_{1}^2 m_{\lambda}^2-2
   m_{\lambda}^4\right) \right. \\ 
 & & \left. +m_{0}^2 \left((\lambda -1)^4 m_{1}^6-(\lambda -1)^2 \left(9 \lambda ^2+8 \lambda +9\right) m_{1}^4 m_{\lambda}^2+\left(-26 \lambda ^2+26 \lambda -9\right)
   m_{1}^2 m_{\lambda}^4+m_{\lambda}^6\right)  \right. \\ 
 & & \left.  +m_{1}^2 m_{\lambda}^2 \left((\lambda -1)^2 m_{1}^2+m_{\lambda}^2\right)^2 \right) \\
  r_3 &=& \frac{1}{2 \lambda   ^3} \left( \lambda ^2 m_{0}^4 \left((2 \lambda -5) m_{1}^2+(3 \lambda -5) m_{\lambda}^2\right)\right. \\ 
 & & \left. +m_{0}^2 \left((\lambda -1)^2 (2 \lambda +3) m_{1}^4+\left(2 \lambda ^3-3 \lambda ^2-3 \lambda
   +2\right) m_{1}^2 m_{\lambda}^2+(3-5 \lambda ) m_{\lambda}^4\right)\right. \\ 
 & & \left. +m_{1}^2 m_{\lambda}^2 \left((\lambda -1)^2 (3 \lambda +2) m_{1}^2+(2-5 \lambda ) m_{\lambda}^2\right) \right)\\
  r_4 &=& \frac{1}{4 \lambda ^2} \left(\lambda ^2 m_{0}^4-2 m_{0}^2 \left(\left(\lambda ^2-\lambda +6\right) m_{1}^2+\left(6 \lambda ^2-11 \lambda +6\right) m_{\lambda}^2\right)+(\lambda -1)^2 m_{1}^4\right. \\ 
 & & \left. -2 \left(6
   \lambda ^2-\lambda +1\right) m_{1}^2 m_{\lambda}^2+m_{\lambda}^4 \right) \\
  r_5 &=& \frac{1}{\lambda } \left( -(\lambda -2) m_0^2-(\lambda +1) m_1^2+(2 \lambda -1) m_{\lambda}^2 \right) \, .
\end{eqnarray*}

The characteristic polynomial for $\mathcal{O}_2$ is 
\begin{equation}
P_{\mathcal{O}_2} (\mathcal{O}_2 ; \lambda , m_{0,1,\lambda}) = \mathcal{O}_2^6 + \sum\limits_{j=0}^5 s_j (\lambda , m_{0,1,\lambda}) \mathcal{O}_2^j
\end{equation}
where
\begin{eqnarray*}
 s_0 &=& \frac{(\lambda ^2-\lambda +1)^{12} m_p^4 m_1^4 m_{\lambda}^4 }{16 (\lambda -1)^{12} \lambda ^{12}} \left(3 (\lambda -1)^4 m_1^4 \left(\lambda ^2m_p^2+m_{\lambda}^2\right)+\left(\lambda ^2   m_p^2+m_{\lambda}^2\right)^3  \right. \\ 
 & & \left.  +3 (\lambda -1)^2 m_1^2 \left(\lambda ^4 m_p^4-7 \lambda ^2 m_p^2 m_{\lambda}^2+m_{\lambda}^4\right)+(\lambda -1)^6 m_1^6\right)^2 
\end{eqnarray*}
\begin{eqnarray*}
 s_1 &=&  \frac{\left(\lambda ^2-\lambda +1\right)^{10} m_p^2 m_1^2 m_{\lambda}^2}{8 (\lambda -1)^{10} \lambda ^{10}} \left(18 m_p^2 m_1^2 m_{\lambda}^2 \left((\lambda -2) \lambda ^4 m_p^4-(\lambda -1)^2   m_1^2 \left((1-2 \lambda ) \lambda ^2 m_p^2+(\lambda -2) m_{\lambda}^2\right)  \right. \right.\\ 
 & & \left.  \left.  -\lambda ^2 (\lambda +1) m_p^2 m_{\lambda}^2+(\lambda -1)^4 (\lambda +1) m_1^4+(1-2   \lambda ) m_{\lambda}^4\right)^2  \right. \\ 
 & & \left.   +\left(3 (\lambda -1)^4 m_1^4 \left(\lambda ^2 m_p^2+m_{\lambda}^2\right)+\left(\lambda ^2 m_p^2+m_{\lambda}^2\right)^3+3 (\lambda -1)^2   m_1^2 \left(\lambda ^4 m_p^4-7 \lambda ^2 m_p^2 m_{\lambda}^2+m_{\lambda}^4\right) \right. \right. \\ 
 & & \left.  \left. +(\lambda -1)^6 m_1^6\right) \left((\lambda -1)^4 m_1^6   \left(m_p^2+m_{\lambda}^2\right)+m_p^2 m_{\lambda}^2 \left(\lambda ^2m_p^2+m_{\lambda}^2\right)^2 \right. \right. \\ 
 & & \left.  \left.  +(\lambda -1)^2 m_1^4 \left(2 \lambda ^2 m_p^4-\left(9 \lambda   ^2+8 \lambda +9\right) m_p^2 m_{\lambda}^2+2 m_{\lambda}^4\right)+m_1^2 \left(\lambda ^4 m_p^6\right. \right.\right. \\ 
 & & \left.  \left.\left.+\lambda ^2 \left(-9 \lambda ^2+26 \lambda -26\right) m_p^4   m_{\lambda}^2+\left(-26 \lambda ^2+26 \lambda -9\right) m_p^2 m_{\lambda}^4+m_{\lambda}^6\right)\right)\right) 
\end{eqnarray*}
\begin{eqnarray*}
 s_2 &=&  \frac{\left(\lambda ^2-\lambda +1\right)^8}{16 (\lambda -1)^8 \lambda ^8} \left(24 m_p^2 m_1^2 m_{\lambda}^2 \left((\lambda -2) \lambda ^4 m_p^4-(\lambda -1)^2 m_1^2 \left((1-2 \lambda ) \lambda ^2   m_p^2+(\lambda -2) m_{\lambda}^2\right)  \right. \right. \\ 
 & & \left.  \left.  -\lambda ^2 (\lambda +1) m_p^2 m_{\lambda}^2+(\lambda -1)^4 (\lambda +1) m_1^4+(1-2 \lambda ) m_{\lambda}^4\right) \left((\lambda -1)^2
   m_1^4 \left((2 \lambda +3) m_p^2+(3 \lambda +2) m_{\lambda}^2\right) \right. \right. \\ 
 & & \left.  \left. +m_p^2 m_{\lambda}^2 \left(\lambda ^2 (3 \lambda -5) m_p^2+(3-5 \lambda )   m_{\lambda}^2\right)+m_1^2 \left(\lambda ^2 (2 \lambda -5) m_p^4+\left(2 \lambda ^3-3 \lambda ^2-3 \lambda +2\right) m_p^2 m_{\lambda}^2 \right. \right. \right.\\ 
 & & \left.  \left.  \left. +(2-5 \lambda )   m_{\lambda}^4\right)\right)+2 m_p^2 m_1^2 m_{\lambda}^2 \left(3 (\lambda -1)^4 m_1^4 \left(\lambda ^2 m_p^2+m_{\lambda}^2\right)+\left(\lambda ^2   m_p^2+m_{\lambda}^2\right)^3 \right. \right. \\ 
 & & \left.  \left.+3 (\lambda -1)^2 m_1^2 \left(\lambda ^4 m_p^4-7 \lambda ^2 m_p^2 m_{\lambda}^2+m_{\lambda}^4\right)+(\lambda -1)^6 m_1^6\right)   \left(\lambda ^2 m_p^4 \right. \right. \\ 
 & & \left.  \left.-2 m_p^2 \left(\left(\lambda ^2-\lambda +6\right) m_1^2+\left(6 \lambda ^2-11 \lambda +6\right) m_{\lambda}^2\right)+(\lambda -1)^2 m_1^4-2
   \left(6 \lambda ^2-\lambda +1\right) m_1^2 m_{\lambda}^2+m_{\lambda}^4\right)  \right. \\ 
 & & \left.  +\left((\lambda -1)^4 m_1^6 \left(m_p^2+m_{\lambda}^2\right)+m_p^2 m_{\lambda}^2 \left(\lambda
   ^2 m_p^2+m_{\lambda}^2\right)^2+(\lambda -1)^2 m_1^4 \left(2 \lambda ^2 m_p^4\right. \right.\right. \\ 
 & & \left.  \left.\left. -\left(9 \lambda ^2+8 \lambda +9\right) m_p^2 m_{\lambda}^2+2
   m_{\lambda}^4\right)+m_1^2 \left(\lambda ^4 m_p^6+\lambda ^2 \left(-9 \lambda ^2+26 \lambda -26\right) m_p^4 m_{\lambda}^2\right. \right.\right. \\ 
 & & \left.  \left.\left.+\left(-26 \lambda ^2+26 \lambda -9\right)
   m_p^2 m_{\lambda}^4+m_{\lambda}^6\right)\right)^2\right)
\end{eqnarray*}
\begin{eqnarray*}
 s_3 &=&  \frac{\left(\lambda ^2-\lambda +1\right)^6}{8 (\lambda -1)^6 \lambda ^6}  \left((\lambda -1)^6 m_1^{10} \left(m_p^2+m_{\lambda}^2\right)+(\lambda -1)^4 m_1^8 \left(\left(8 \lambda ^2+26 \lambda +6\right)   m_p^4  \right. \right.\\ 
 & & \left. \left. +\left(21 \lambda ^2+104 \lambda +21\right) m_p^2 m_{\lambda}^2+2 \left(3 \lambda ^2+13 \lambda +4\right) m_{\lambda}^4\right)+(\lambda -1)^2 m_1^6 \left(\lambda ^2   \left(14 \lambda ^2-14 \lambda -83\right) m_p^6\right. \right. \\ 
 & & \left.  \left.  +\left(74 \lambda ^4+4 \lambda ^3-99 \lambda ^2+176 \lambda +72\right) m_p^4 m_{\lambda}^2+\left(72 \lambda ^4+176 \lambda ^3-99
   \lambda ^2+4 \lambda +74\right) m_p^2 m_{\lambda}^4\right. \right. \\ 
 & & \left.  \left.  +\left(-83 \lambda ^2-14 \lambda +14\right) m_{\lambda}^6\right)+m_1^4 \left(2 \lambda ^4 \left(4 \lambda ^2-21 \lambda
   +20\right) m_p^8\right. \right. \\ 
 & & \left.  \left.  +\lambda ^2 \left(74 \lambda ^4-300 \lambda ^3+357 \lambda ^2-286 \lambda +227\right) m_p^6 m_{\lambda}^2\right. \right. \\ 
 & & \left.  \left.  +2 \left(82 \lambda ^6-246 \lambda ^5+249 \lambda ^4-88   \lambda ^3+249 \lambda ^2-246 \lambda +82\right) m_p^4 m_{\lambda}^4\right. \right. \\ 
 & & \left.  \left.  +\left(227 \lambda ^4-286 \lambda ^3+357 \lambda ^2-300 \lambda +74\right) m_p^2 m_{\lambda}^6+2 \left(20
   \lambda ^2-21 \lambda +4\right) m_{\lambda}^8\right) \right. \\ 
 & &  \left.  +m_p^2 m_{\lambda}^2 \left(\lambda ^6 m_p^8+2 \lambda ^4 \left(3 \lambda ^2-19 \lambda +20\right) m_p^6
   m_{\lambda}^2+\lambda ^2 \left(-83 \lambda ^2+180 \lambda -83\right) m_p^4 m_{\lambda}^4\right. \right. \\ 
 & & \left.  \left.  +2 \left(20 \lambda ^2-19 \lambda +3\right) m_p^2
   m_{\lambda}^6+m_{\lambda}^8\right)+m_1^2 \left(\lambda ^6 m_p^{10}+\lambda ^4 \left(21 \lambda ^2-146 \lambda +146\right) m_p^8 m_{\lambda}^2\right. \right. \\ 
 & & \left.  \left.  +\lambda ^2 \left(72 \lambda   ^4-464 \lambda ^3+861 \lambda ^2-622 \lambda +227\right) m_p^6 m_{\lambda}^4+\left(227 \lambda ^4-622 \lambda ^3+861 \lambda ^2-464 \lambda +72\right) m_p^4   m_{\lambda}^6\right. \right. \\ 
 & & \left.  \left.  +\left(146 \lambda ^2-146 \lambda +21\right) m_p^2 m_{\lambda}^8+m_{\lambda}^{10}\right)\right)
\end{eqnarray*}
\begin{eqnarray*}
 s_4 &=&  \frac{\left(\lambda ^2-\lambda +1\right)^4}{16 (\lambda -1)^4 \lambda ^4} \left(\lambda ^4 m_p^8+4 \lambda ^2 \left(4 \lambda ^2-33 \lambda +34\right) m_p^6 m_{\lambda}^2+2 \left(24 \lambda ^4-160 \lambda
   ^3+299 \lambda ^2-160 \lambda +24\right) m_p^4 m_{\lambda}^4 \right. \\ 
 & & \left. +4 (\lambda -1)^2 m_1^6 \left(\left(5 \lambda ^2+25 \lambda +4\right) m_p^2+\left(4 \lambda ^2+25 \lambda
   +5\right) m_{\lambda}^2\right)+4 \left(34 \lambda ^2-33 \lambda +4\right) m_p^2 m_{\lambda}^6 \right. \\ 
 & & \left. +2 m_1^4 \left(\left(27 \lambda ^4-54 \lambda ^3-37 \lambda ^2+64 \lambda +24\right)
   m_p^4+2 \left(28 \lambda ^4-39 \lambda ^3+78 \lambda ^2-39 \lambda +28\right) m_p^2 m_{\lambda}^2 \right.  \right. \\ 
 & & \left. \left.  +\left(24 \lambda ^4+64 \lambda ^3-37 \lambda ^2-54 \lambda +27\right)
   m_{\lambda}^4\right)+4 m_1^2 \left(\lambda ^2 \left(5 \lambda ^2-35 \lambda +34\right) m_p^6\right. \right. \\ 
 & & \left.  \left.   +\left(28 \lambda ^4-73 \lambda ^3+129 \lambda ^2-112 \lambda +56\right) m_p^4
   m_{\lambda}^2+\left(56 \lambda ^4-112 \lambda ^3+129 \lambda ^2-73 \lambda +28\right) m_p^2 m_{\lambda}^4\right. \right. \\ 
 & & \left.  \left.  +\left(34 \lambda ^2-35 \lambda +5\right) m_{\lambda}^6\right)+(\lambda -1)^4   m_1^8+m_{\lambda}^8\right)
\end{eqnarray*}
\begin{eqnarray*}
  s_5 &=& \frac{\left(\lambda ^2-\lambda +1\right)^2 }{2 (\lambda -1)^2 \lambda ^2} \left(\left(\lambda ^2-8 \lambda +8\right) m_p^4+m_p^2 \left(\left(6 \lambda ^2-6 \lambda +4\right) m_1^2+2 \left(2 \lambda   ^2-\lambda +2\right) m_{\lambda}^2\right) \right. \\ 
 & & \left. +\left(\lambda ^2+6 \lambda +1\right) m_1^4+2 \left(2 \lambda ^2-3 \lambda +3\right) m_1^2 m_{\lambda}^2+\left(8 \lambda ^2-8 \lambda
   +1\right) m_{\lambda}^4\right) \, .
\end{eqnarray*}

\bibliographystyle{JHEP}

\end{document}